\documentclass[epj,final]{svjour}
\smartqed  
\RequirePackage{graphicx}
\begin{document}

\title{Hadron transverse momentum distributions of the Tsallis normalized and unnormalized statistics}

\author{A.S.~Parvan\inst{1}\inst{2}\inst{3}, T.~Bhattacharyya\inst{1}}

\institute{\inst{1}Bogoliubov Laboratory of Theoretical Physics, Joint Institute for Nuclear Research, Dubna, Russia \\ \inst{2}Department of Theoretical Physics, Horia Hulubei National Institute of Physics and Nuclear Engineering, Bucharest-Magurele, Romania \\ \inst{3}Institute of Applied Physics, Moldova Academy of Sciences, Chisinau, Republic of Moldova}

\date{Received: date / Revised version: date}

\abstract{The exact analytical formulas for the transverse momentum distributions of the Bose-Einstein, Fermi-Dirac and Maxwell-Boltzmann statistics of particles with nonzero mass in the framework of the Tsallis normalized and Tsallis unnormalized (also known as Tsallis-1 and Tsallis-2) statistics have been consistently derived. The final exact results were expressed in terms of the series expansions in the integral representation. The zeroth term approximation to both quantum and classical statistics of particles has been introduced. We have revealed that the phenomenological classical Tsallis distribution (widely used in high energy physics) is equal to the distribution of the Tsallis unnormalized statistics in the zeroth term approximation, but the phenomenological quantum Tsallis distributions (introduced by definition on the basis of the generalized entropy of the ideal gas) do not correspond to the distributions of the Tsallis statistics. We have found that in the ranges of the entropic parameter relevant to the processes of high-energy physics ($q<1$ for Tsallis-1 and $q>1$ for Tsallis-2) the Tsallis statistics is divergent. Therefore, to obtain physical results, we have regularized the Tsallis statistics by introducing an upper cut-off in the series expansion. The exact numerical results for the Bose-Einstein, Fermi-Dirac and Maxwell-Boltzmann statistics of particles in the Tsallis normalized and unnormalized statistics have been obtained. We observed that the exact results of the Tsallis statistics strongly enhanced the production of high-$p_{T}$ hadrons in comparison with the usual phenomenological Tsallis distribution function at the same values of $q$. The $q$-duality of the Tsallis normalized and unnormalized statistics for the massive particles was studied.
} 
\titlerunning{Hadron transverse momentum distributions of}
\authorrunning{A.S.~Parvan, T.~Bhattacharyya}


\maketitle

\section{Introduction}\label{sec1}
Power-law functions are used to describe the experimental data for the transverse momentum distribution of particles produced in proton-proton and heavy-ion collisions at the LHC and RHIC energies~\cite{STAR,PHENIX1,ALICE_charged,ALICE_piplus,ALICE_PbPb,CMS1,ATLAS,ALICE_deuteron}. Now the phenomenological transverse momentum distribution~\cite{Cleymans2012,Bediaga00,Beck00} inspired by the Tsallis statistics~\cite{Tsal88} has gained much attention and it is successfully used for the description of the experimental data on high-energy proton-proton reactions~\cite{Rybczynski14,Cleymans13,Azmi14,Cleymans12a,Marques13,Li14,Parvan14,Parvan17a,Biyajima06,Marques15,TsallisTaylor,TsallisRAA,BhattaCleMog,bcmmp} and relativistic heavy-ion collisions~\cite{Shen2019,Shen2019a,Shen2019b,Chaturvedi2018,Gao2016,Thakur2017}. However, it has some dificulties in relation to the fundamental principles of thermodynamics and statistical mechanics. Recently, it was analytically proved that this phenomenological Tsallis-like distribution in the case of the massless particles corresponds to the zeroth term approximation of the improperly normalized variant of the Tsallis statistics, namely the Tsallis-2 statistics (see Ref.~\cite{Parvan17}). In the present paper the same statement for the phenomenological Tsallis-like distribution is proved in the case of massive particles.

Hadronic transverse momentum ($p_{T}$) spectra carry essential information about the hadron production mechanism in high-energy collision events. It is one of the main observables measured in high-energy proton-proton ($p-p$) and nucleus-nucleus ($A-A$) collisions which allows to determine many other experimental quantities like nuclear modification factors~\cite{TsallisRAA,Acharya19}, flows~\cite{Yagi05}, etc. The hadronic transverse momentum spectra give us information about the freeze-out parameters like temperature and chemical potential~\cite{Thakur16}. Ratios of the hadron yields which are obtained by integration of the transverse momentum and rapidity distributions provide ample information about the properties of the hadronic matter and the quark-gluon plasma (QGP), the state of strongly interacting matter probably formed at the very high temperatures and energy densities~\cite{Adam17,Adam16,Acharya19a,Alt08,Abelev10,Bearden05}.

In the present paper, we consider the original formulation of the Tsallis statistics~\cite{Tsal88,Tsal98}, that is correctly connected with the first principles of equilibrium thermodynamics. In particular, the equilibrium probabilities of microstates are obtained from the second law of thermodynamics (Jaynes principle or maximum entropy principle)~\cite{Jaynes2}. There are several classification schemes of the Tsallis statistics depending on the choice of the definition of the expectation values of the extensive thermodynamic quantities~\cite{Tsal98}. The present paper discusses two of them a) the Tsallis normalized and b) the Tsallis unnormalized statistics, which are also termed as the Tsallis-1 and the Tsallis-2 statistics, respectively \cite{Tsal98}. Generally, the Tsallis-2 statistics is incorrectly defined. In this statistics, the definition of the expectation values of the thermodynamic quantities is not consistent with the norm equation of the probability distribution of microstates. This leads to the incorrect connection of the Tsallis-2 statistics with the second law of thermodynamics.

The current research represents a generalization of an earlier work done by A.S.~Parvan~\cite{Parvan17} for the Maxwell-Boltzmann statistics of particles in the ultrarelativistic limit where particle mass was neglected and analytical results were obtained. In the present paper, we consider all the three particle distributions with nonzero mass -- Maxwell-Boltzmann, Fermi-Dirac and Bose-Einstein in the framework of the Tsallis normalized and unnormalized statistics in the grand canonical ensemble and provide a consistent derivation of the corresponding transverse momentum spectra which can be solved numerically. The analytical results for the integrals of the relativistic distribution of the Tsallis statistics can be obtained only in the case of the Maxwell-Boltzmann statistics of massless particles~\cite{Parvan17}. For the massive particles the analytical formulas can not be obtained and the integrals can be solved only numerically. The  ultrarelativistic Tsallis distribution function of massless particles describes the experimental data for the transverse momentum distributions of hadrons created in high energy collisions very well~\cite{Parvan16}. Still, the ultrarelativistic Tsallis distribution slightly disagrees with the experimental data at low values of $p_{T}$ at low energies of the NA61/SHINE Collaboration~\cite{Parvan16}. However, the consideration of the mass of particles in the Tsallis distributions allows to describe correctly the low values of the transverse momentum $p_{T}$ of light hadrons (pions) at low collision energies (cf. the results, for example, of Refs.~\cite{Parvan16} and~\cite{Parvan17a}, and the results of the fit given in the present paper). The mass of the particles in the Tsallis formalism is also important for heavy hadrons. We are interested in the exact approach to the transverse momentum distribution of the massive hadrons coming out of the high-energy collision experiments using the Tsallis normalized statistics formalism. The exact transverse momentum distribution of the massive hadrons in the framework of the improper Tsallis-2 formalism is needed only to find the origin of the phenomenological Tsallis-like distribution.

The exact approach of the Tsallis nonextensive statistics has already been applied to describe experimental data on the transverse momentum distribution of hadrons created in high-energy proton-proton collisions in Ref.~\cite{Parvan16}. These results were obtained for the case of the Maxwell-Boltzmann statistics of massless particles in the ultrarelativistic limit and in the method in which the cut-off parameter of the series expansion was found from the inflection point. In the present paper, the cut-off parameter for the particles with nonzero mass was calculated from the minima of the characteristic function. Both of these methods were discussed and compared in Ref.~\cite{ParvanBaldin}. In Ref.~\cite{Parvan17}, it was demonstrated that in the ultrarelativistic limit the phenomenological Tsallis distribution for the Maxwell-Boltzmann statistics of particles proposed in Ref.~\cite{Cleymans2012} exactly coincides with the transverse momentum distribution of the Tsallis unnormalized (Tsallis-2) statistics in the zeroth term approximation. It was also proved that this distribution in the case of massless particles can be obtained from the zeroth term approximation of the Tsallis normalized statistics by the transformation of the entropic parameter $q\to 1/q$. In the present study, we obtain the same results for the particles with nonzero mass.

The Tsallis nonextensive statistics has been applied to the quantum systems. The exact solutions to the quantum ideal gas in the grand canonical ensemble were obtained in Refs.~\cite{Rajagopal98,Lenzi99} by using the integral representation approach based on the Hilhorst formula~\cite{Abramowitz,Prato}. The exact results for the quantum systems in the framework of the Tsallis statistics defined in the optimal Lagrange multiplier method were obtained by Hasegawa in~\cite{Hasegawa09}. The evaluation of an integral was performed numerically or in an approximate way~\cite{Rajagopal98,Lenzi99,Hasegawa09,Aragao03}. In the present research, we use the power series expansion of the exponential function containing the Boltzmann-Gibbs grand-canonical thermodynamic potential of the system. This method was introduced in Refs.~\cite{Parvan17,Parvan2015,Parvan16}. It provides exact results. In the relativistic case the integrals in each term of the series can be evaluated analytically only for the Maxwell-Boltzmann massless ideal gas~\cite{Parvan17,Parvan16}. However, for the Bose-Einstein, Fermi-Dirac and Maxwell-Boltzmann statistics of particles with nonzero mass they may not be solved analytically and can be evaluated only numerically.

The Tsallis statistics is confronted with the problem of the physical interpretation of $q$ entropic parameter. In Ref.~\cite{Wilk2000}, it was shown that the parameter $q$ of the L\'evi probability distribution or the phenomenological single-particle Tsallis-like distribution is given by the fluctuations of the inverse temperature of the usual exponential distribution. The scaling properties of the variable $z=1/(q-1)$ and their relations to the thermodynamic limit in the Tsallis statistics were found in Refs.~\cite{Botet2,Botet1}. In Refs.~\cite{Parv2,Parv2a,Parvan2015}, it was shown that if the entropic variable $z$ is an extensive thermodynamic variable of state then the Tsallis statistics is thermodynamically self - consistent in the thermodynamic limit, i.e., it agrees with the requirements of the equilibrium thermodynamics. In Ref.~\cite{Wilk2014}, the authors proposed the possibility of complex $q$ in the phenomenological single-particle Tsallis-like distribution to explain the oscillation of the ratio of data/fit observed in experimental data on the transverse momentum distribution of hadrons created in $pp$ collisions at high energies. This procedure can be certainly implemented for the single-particle Tsallis-like functions defined as in Ref.~\cite{Wilk2014} which supposes that the phenomenological single-particle Tsallis-like distribution has the same form as the Tsallis distribution of microstates of a system. However, this method may not be valid for the case of the Tsallis statistics~\cite{Tsal88} derived from the second law of thermodynamics because the exact single-particle distribution functions of the Tsallis statistics do not preserve the form of the Tsallis distribution of microstates of system (see Ref.~\cite{Parvan17} and calculations given below).

The paper is organized as follows:  in the next section, we discuss the general formalism of the Tsallis-1 statistics and derive the classical and quantum transverse momentum distributions. In Sec.~\ref{sec3}, we repeat the same calculations for the Tsallis-2 statistics. Sec.~\ref{sec4} discusses the results for the transverse momentum spectra of hadrons. We summarize and conclude in Sec.~\ref{sec5}.

\section{Tsallis normalized statistics in the grand canonical ensemble}\label{sec2}
\subsection{General formalism}
The Tsallis normalized statistics or the Tsallis-$1$ statistics~\cite{Tsal88} is defined by the generalized entropy with the probabilities $p_{i}$ of the microstates of the system normalized to unity~\cite{Tsal88,Tsal98}
\begin{equation}\label{1}
    S = \sum\limits_{i} \frac{p_{i}^{q}-p_{i}}{1-q}, \qquad  \sum\limits_{i} p_{i}=1
\end{equation}
and by the standard expectation values
\begin{equation}\label{2}
   \langle A \rangle = \sum\limits_{i} p_{i} A_{i},
\end{equation}
where $q\in\mathbf{R}$ is a real parameter taking values $0<q<\infty$. Here and throughout the paper we use the system of natural units $\hbar=c=k_{B}=1$. Note that in the Gibbs limit $q\to 1$, the entropy (\ref{1}) recovers the Boltzmann-Gibbs-Shannon entropy, $S=-\sum_{i} p_{i} \ln p_{i}$, and the Tsallis normalized statistics is reduced to the Boltzmann-Gibbs statistics.

The thermodynamic potential $\Omega$ of the grand canonical ensemble is the Legendre transform of the fundamental thermodynamic potential $\langle H \rangle$ and it can be written as~\cite{Parvan2015}
\begin{eqnarray}\label{3}
 \Omega &=& \langle H \rangle -TS-\mu \langle N \rangle \nonumber \\
  &=&  \sum\limits_{i}  p_{i} \left[E_{i}-\mu N_{i} - T \frac{p_{i}^{q-1}-1}{1-q}\right],
\end{eqnarray}
where $\langle H \rangle=\sum_{i}  p_{i} E_{i}$ is the mean energy of the system, $\langle N \rangle=\sum_{i}  p_{i} N_{i}$ is the mean number of particles, and $E_{i}$ and $N_{i}$ are the energy and number of particles, respectively, in the $i$-th microscopic state of the system.

The unknown probabilities $\{p_{i}\}$, which  are constrained by an additional function
\begin{equation}\label{4}
    \varphi=\sum\limits_{i} p_{i} - 1 = 0,
\end{equation}
are obtained from the second law of thermodynamics (the principle of maximum entropy). This means that in the grand canonical ensemble the set of equilibrium probabilities $\{p_{i}\}$ can be found from the constrained local extrema of the thermodynamic potential (\ref{3}) by the method of the Lagrange multipliers (see, for example, Refs.~\cite{Jaynes2,Parvan2015,Krasnov}):
\begin{eqnarray}\label{5}
 \Phi &=& \Omega - \lambda \varphi,  \\ \label{6}
  \frac{\partial \Phi}{\partial p_{i}} &=& 0,
\end{eqnarray}
where $\lambda$ is an arbitrary real constant. Substituting Eqs.~(\ref{3}) and (\ref{4}) into Eqs.~(\ref{5}), (\ref{6}) and using again Eq.~(\ref{4}), we obtain the normalized equilibrium probabilities of the grand canonical ensemble of the Tsallis normalized statistics as ~\cite{Parvan2015}
\begin{equation}\label{7}
p_{i} = \left[1+\frac{q-1}{q}\frac{\Lambda-E_{i}+\mu N_{i}}{T}\right]^{\frac{1}{q-1}}
\end{equation}
and
\begin{equation}\label{8}
    \sum\limits_{i} \left[1+\frac{q-1}{q}\frac{\Lambda-E_{i}+\mu N_{i}}{T}\right]^{\frac{1}{q-1}}=1,
\end{equation}
where $\Lambda\equiv \lambda-T$ and $\partial E_{i}/\partial p_{i}=\partial N_{i}/\partial p_{i}=0$. In the Gibbs limit $q\to 1$, the probability $p_{i}=\exp[(\Lambda-E_{i}+\mu N_{i})/T]$, where $\Lambda=-T\ln Z$ is the thermodynamic potential of grand canonical ensemble and $Z=\sum_{i} \exp[-(E_{i} - \mu N_{i})/T]$ is the partition function.

Substituting Eq.~(\ref{7}) into Eq.~(\ref{2}), we obtain the statistical averages of the Tsallis normalized statistics in the grand canonical ensemble as~\cite{Parvan2015}
\begin{equation}\label{9}
   \langle A \rangle = \sum\limits_{i} A_{i} \left[1+\frac{q-1}{q}\frac{\Lambda-E_{i}+\mu N_{i}}{T}\right]^{\frac{1}{q-1}},
\end{equation}
where the norm function $\Lambda$ is the solution of Eq.~(\ref{8}).

Let us rewrite the probabilities of microstates (\ref{7}), the norm equation (\ref{8}) and the statistical averages (\ref{9}) in the integral representation. To obtain this, we use the formulas for the integral representation of the Gamma - function~\cite{Abramowitz,Prato}:
\begin{eqnarray}\label{10}
  x^{-y} &=& \frac{1}{\Gamma(y)} \int\limits_{0}^{\infty}  t^{y-1} e^{-tx}  dt,   \mathrm{Re}(x)>0,  \mathrm{Re}(y)>0, \\ \label{11}
   x^{y-1} &=& \Gamma(y) \frac{i}{2\pi} \oint\limits_{C} (-t)^{-y} e^{-tx}  dt,   \mathrm{Re}(x)>0, |y|<\infty. \;\;\;\;\;\;
\end{eqnarray}
Using Eqs.~(\ref{10}) and (\ref{11}) for $q<1$ and $q>1$, respectively, we obtain the formulas for the integral representation of the probabilities of microstates (\ref{7}) as
\begin{equation}\label{12}
  p_{i}=\frac{1}{\Gamma\left(\frac{1}{1-q}\right)} \int\limits_{0}^{\infty} t^{\frac{q}{1-q}} e^{-t\left[1+\frac{q-1}{q}\frac{\Lambda-E_{i}+\mu N_{i}}{T}\right]} dt \quad \mathrm{for} \quad q<1
\end{equation}
and
\begin{eqnarray}\label{13}
  p_{i}&=&\Gamma\left(\frac{q}{q-1}\right)  \frac{i}{2\pi} \oint\limits_{C} (-t)^{\frac{q}{1-q}} e^{-t\left[1+\frac{q-1}{q}\frac{\Lambda-E_{i}+\mu N_{i}}{T}\right]} dt \nonumber \\
  && \mathrm{for} \quad q>1.
\end{eqnarray}
These equations link the probability distribution of the Tsallis normalized statistics with the probability distribution of the Boltzmann-Gibbs statistics. The norm equation (\ref{8}) in the integral representation can be rewritten as~\cite{ParvanBaldin}
\begin{equation}\label{14}
 \frac{1}{\Gamma\left(\frac{1}{1-q}\right)} \int\limits_{0}^{\infty} t^{\frac{q}{1-q}} e^{-t\left[1+\frac{q-1}{q}\frac{\Lambda-\Omega_{G}\left(\beta'\right)}{T}\right]} dt =1 \quad \mathrm{for} \quad q<1
\end{equation}
and
\begin{eqnarray}\label{15}
  && \Gamma\left(\frac{q}{q-1}\right)  \frac{i}{2\pi} \oint\limits_{C} (-t)^{\frac{q}{1-q}} e^{-t\left[1+\frac{q-1}{q}\frac{\Lambda-\Omega_{G}\left(\beta'\right)}{T}\right]} dt =1 \nonumber \\ && \mathrm{for} \quad q>1,
\end{eqnarray}
where $\beta'=t(1-q)/qT$ and
\begin{eqnarray}\label{16}
  \Omega_{G}\left(\beta'\right) &=& -\frac{1}{\beta'} \ln Z_{G}\left(\beta'\right), \\ \label{16a}  Z_{G}\left(\beta'\right) &=& \sum\limits_{i} e^{-\beta'(E_{i}-\mu N_{i})}.
\end{eqnarray}
Equations~(\ref{15}) and (\ref{16}) link the Tsallis norm function $\Lambda$ with the thermodynamic potential of the Boltzmann-Gibbs statistics. The statistical averages (\ref{9}) can also be written in the integral representation. Using Eqs.~(\ref{10}) and (\ref{11}) for $q<1$ and $q>1$, respectively, we obtain~\cite{ParvanBaldin}
\begin{eqnarray}\label{17}
  \langle A \rangle &=& \frac{1}{\Gamma\left(\frac{1}{1-q}\right)} \int\limits_{0}^{\infty}  t^{\frac{q}{1-q}} e^{-t\left[1+\frac{q-1}{q}\frac{\Lambda-\Omega_{G}\left(\beta'\right)}{T}\right]} \langle A \rangle_{G}\left(\beta'\right) dt   \nonumber \\ && \mathrm{for} \quad q<1
\end{eqnarray}
and
\begin{eqnarray}\label{18}
   \langle A \rangle &=& \Gamma\left(\frac{q}{q-1}\right)  \frac{i}{2\pi}  \oint\limits_{C}  (-t)^{\frac{q}{1-q}} \nonumber \\ && e^{-t\left[1+\frac{q-1}{q}\frac{\Lambda-\Omega_{G}\left(\beta'\right)}{T}\right]}  \langle A \rangle_{G}\left(\beta'\right) dt  \quad \mathrm{for}\quad q>1, \;\;\;\;\;\;
\end{eqnarray}
where
\begin{equation}\label{19}
  \langle A \rangle_{G}\left(\beta'\right) =\frac{1}{Z_{G}\left(\beta'\right)} \sum\limits_{i} A_{i} e^{-\beta'(E_{i}-\mu N_{i})}.
\end{equation}
Equations~(\ref{17}) and (\ref{18}) link the statistical averages of the Tsallis normalized statistics with the corresponding statistical averages (\ref{19})  of the Boltzmann-Gibbs statistics.

\subsection{Transverse momentum distribution in the Tsallis-1 statistics}

\subsubsection{Exact results}
\paragraph{General case}
The transverse momentum distribution in the grand canonical ensemble in the Tsallis normalized statistics is derived from Eqs.~(\ref{17})--(\ref{19}). For the ideal gas with the Fermi-Dirac $(\eta=1)$, Bose-Einstein $(\eta=-1)$ and Maxwell-Boltzmann $(\eta=0)$ statistics of particles, the transverse momentum distribution can be written in the form of the series expansion as (see Appendix~\ref{ap1})
\begin{eqnarray}\label{36}
  \frac{d^{2}N}{dp_{T}dy} &=& \frac{gV}{(2\pi)^{2}} p_{T}  m_{T} \cosh y  \sum\limits_{n=0}^{\infty}  \frac{1}{n!\Gamma\left(\frac{1}{1-q}\right)} \nonumber \\ &&
  \int\limits_{0}^{\infty} t^{\frac{q}{1-q}} e^{-t\left[1+\frac{q-1}{q}\frac{\Lambda}{T}\right]}    \frac{\left(-\beta'\Omega_{G}\left(\beta'\right)\right)^{n}}{e^{\beta' (m_{T} \cosh y-\mu)}+\eta} dt \nonumber \\ &&   \quad \mathrm{for} \quad q<1
\end{eqnarray}
and
\begin{eqnarray}\label{37}
 \frac{d^{2}N}{dp_{T}dy} &=& \frac{gV}{(2\pi)^{2}} p_{T}  m_{T} \cosh y \sum\limits_{n=0}^{\infty} \frac{\Gamma\left(\frac{q}{q-1}\right)}{n!}    \frac{i}{2\pi} \nonumber \\ &&
 \oint\limits_{C} (-t)^{\frac{q}{1-q}} e^{-t\left[1+\frac{q-1}{q}\frac{\Lambda}{T}\right]} \frac{\left(-\beta'\Omega_{G}\left(\beta'\right)\right)^{n}}{e^{\beta' (m_{T} \cosh y-\mu)}+\eta} dt
 \nonumber \\ &&  \quad \mathrm{for} \quad q>1, \;\;\;\;\;\;
\end{eqnarray}
where $p_{T}$ and $y$ are the transverse momentum and rapidity variables, respectively, and $m_{T}=\sqrt{p_{T}^{2}+m^{2}}$ is the transverse mass of particle.

The norm function $\Lambda$ is calculated from the norm equation~(\ref{8}) which in the grand canonical ensemble for both quantum and classical statistics of particles in the integral representation (\ref{14}) and (\ref{15}) can be rewritten in the form of a series expansion as (see Appendix~\ref{ap1})
\begin{eqnarray}\label{21}
 && \sum\limits_{n=0}^{\infty} \frac{1}{n!\Gamma\left(\frac{1}{1-q}\right)} \int\limits_{0}^{\infty} t^{\frac{q}{1-q}} e^{-t\left[1+\frac{q-1}{q}\frac{\Lambda}{T}\right]} \left(-\beta'\Omega_{G}\left(\beta'\right)\right)^{n} dt =1 \nonumber \\  && \mathrm{for} \quad q<1
\end{eqnarray}
and
\begin{eqnarray}\label{22}
  \sum\limits_{n=0}^{\infty} && \frac{\Gamma\left(\frac{q}{q-1}\right)}{n!}    \frac{i}{2\pi} \oint\limits_{C} dt (-t)^{\frac{q}{1-q}} \nonumber \\  && e^{-t\left[1+\frac{q-1}{q}\frac{\Lambda}{T}\right]} \left(-\beta'\Omega_{G}\left(\beta'\right)\right)^{n}  =1 \quad \mathrm{for} \quad q>1,
\end{eqnarray}
where
\begin{equation}\label{23}
  -\beta'\Omega_{G}\left(\beta'\right)= \sum\limits_{\mathbf{p},\sigma} \ln \left[1+\eta e^{-\beta' (\varepsilon_{\mathbf{p}}-\mu)} \right]^{\frac{1}{\eta}}
\end{equation}
is the thermodynamic potential for the ideal gas of the Boltzmann-Gibbs statistics of microstates in which $\varepsilon_{\mathbf{p}}=\sqrt{\mathbf{p}^2+m^{2}}$ is one-particle energy and $m$ is mass of a particle.

\paragraph{Maxwell-Boltzmann statistics of particles}
In the case of the Maxwell-Boltzmann statistics of particles in the limit $\eta\to 0$, the thermodynamic potential of the ideal gas of the Boltzmann-Gibbs statistics (\ref{23}) takes the following form
\begin{equation}\label{38}
  \Omega_{G}\left(\beta'\right)= - \frac{gV}{2\pi^{2}} \frac{m^{2}}{\beta'^{2}} e^{\beta'\mu} K_{2}\left(\beta'm\right),
\end{equation}
where $K_{\nu}(z)$ is the modified Bessel function of the second kind. Substituting Eq.~(\ref{38}) into Eqs.~(\ref{21}) and (\ref{22}) and taking the limit $\eta=0$, we obtain the norm equation for the Maxwell-Boltzmann statistics of particles as
\begin{eqnarray}
 && \sum\limits_{n=0}^{\infty} \frac{\omega^{n}}{n!} \frac{1}{\Gamma\left(\frac{1}{1-q}\right)} \int\limits_{0}^{\infty} t^{\frac{q}{1-q}-n} e^{-t\left[1+\frac{q-1}{q}\frac{\Lambda+\mu n}{T}\right]} \nonumber \\ && \qquad \left(K_{2}\left(\frac{t(1-q)m}{qT} \right)\right)^{n} dt = 1 \quad \mathrm{for} \quad q<1  \nonumber
\end{eqnarray}
which can be briefly written as
\begin{equation}\label{39}
 \sum\limits_{n=0}^{\infty} \phi(n) = 1
\end{equation}
and
\begin{eqnarray}\label{40}
 && \sum\limits_{n=0}^{\infty} \frac{(-\omega)^{n}}{n!}  \Gamma\left(\frac{q}{q-1}\right)  \frac{i}{2\pi} \oint\limits_{C} (-t)^{\frac{q}{1-q}-n} e^{-t\left[1+\frac{q-1}{q}\frac{\Lambda+\mu n}{T}\right]} \nonumber \\ && \qquad  \left(K_{2}\left(\frac{t(1-q)m}{qT} \right)\right)^{n} dt =1 \quad \mathrm{for} \quad q>1,
\end{eqnarray}
where
\begin{equation}\label{41}
  \omega=\frac{gVTm^{2}}{2\pi^{2}} \frac{q}{1-q}.
\end{equation}
Note that in the ultrarelativistic limit $(m=0)$, Eqs.~(\ref{39}) and (\ref{40}) recover Eq.~(18) of Ref.~\cite{Parvan17}.

Substituting Eq.~(\ref{38}) into Eqs.~(\ref{36}) and (\ref{37}) and taking the limit $\eta=0$, we obtain the transverse momentum distribution for the Maxwell-Boltzmann statistics of particles as

\begin{eqnarray}\label{50}
  \frac{d^{2}N}{dp_{T}dy} &=& \frac{gV}{(2\pi)^{2}} p_{T}  m_{T} \cosh y  \sum\limits_{n=0}^{\infty} \frac{\omega^{n}}{n!} \frac{1}{\Gamma\left(\frac{1}{1-q}\right)} \nonumber \\
   &\times&   \int\limits_{0}^{\infty} t^{\frac{q}{1-q}-n}
 e^{-t\left[1+\frac{q-1}{q}\frac{\Lambda-m_{T} \cosh y+\mu(n+1)}{T}\right]} \nonumber \\ &\times& \left(K_{2}\left(\frac{t(1-q)m}{qT} \right)\right)^{n} dt \quad \mathrm{for} \quad q<1
\end{eqnarray}
and
\begin{eqnarray}\label{51}
 \frac{d^{2}N}{dp_{T}dy} &=& \frac{gV}{(2\pi)^{2}} p_{T}  m_{T} \cosh y \sum\limits_{n=0}^{\infty} \frac{(-\omega)^{n}}{n!}  \Gamma\left(\frac{q}{q-1}\right) \nonumber \\
   &\times& \frac{i}{2\pi} \oint\limits_{C} (-t)^{\frac{q}{1-q}-n} e^{-t\left[1+\frac{q-1}{q}\frac{\Lambda-m_{T} \cosh y+\mu(n+1)}{T}\right]} \nonumber \\ &\times& \left(K_{2}\left(\frac{t(1-q)m}{qT} \right)\right)^{n} dt \quad \mathrm{for} \quad q>1.
\end{eqnarray}
In the ultrarelativistic limit $(m=0)$, Eqs.~(\ref{50}) and (\ref{51}) recover Eq.~(34) of Ref.~\cite{Parvan17}. See Appendix~\ref{ap1} for more detail.

\subsubsection{Zeroth term approximation}
Let us find the transverse momentum distribution given above in the zeroth term approximation introduced in Ref.~\cite{Parvan17}, i.e. we retain only the zeroth term $(n=0)$ in the series expansion. Taking $n=0$ in Eqs.~(\ref{21}) and (\ref{22}) and using Eqs.~(\ref{10}) and (\ref{11}), we obtain that the norm function $\Lambda=0$. Substituting this value of $\Lambda$ into Eqs.~(\ref{36}) and (\ref{37}) and considering only the zeroth term and the equation
\begin{equation}\label{54}
\frac{1}{e^{x}+\eta} = \sum\limits_{k=0}^{\infty} (-\eta)^{k} e^{-x(k+1)},
\end{equation}
where $|e^{-x}|<1$, and using Eqs.~(\ref{10}) and (\ref{11}), we obtain the transverse momentum distribution in the zeroth term approximation as
\begin{eqnarray}\label{65}
\frac{d^{2}N}{dp_{T}dy} &=& \frac{gV}{(2\pi)^{2}} p_{T}  m_{T} \cosh y \nonumber \\ &\times& \sum\limits_{k=0}^{\infty} (-\eta)^{k}  \left[1+(k+1) \frac{1-q}{q} \frac{m_{T} \cosh y-\mu}{T} \right]^{\frac{1}{q-1}}  \nonumber \\ &&  \quad \mathrm{for} \quad \eta=-1,0,1   \;\;\;\;\;\;\;\;\;\;
\end{eqnarray}
or
\begin{eqnarray}\label{66}
\frac{d^{2}N}{dp_{T}dy} &=& \frac{gV}{(2\pi)^{2}} p_{T}  m_{T} \cosh y  \nonumber \\ &\times& \left[1+ \frac{1-q}{q} \frac{m_{T} \cosh y-\mu}{T} \right]^{\frac{1}{q-1}} \nonumber \\ &&  \quad \mathrm{for} \quad \eta=0.
\end{eqnarray}
In the ultrarelativistic limit $(m=0)$, Eq.~(\ref{66}) recovers Eq.~(44) of Ref.~\cite{Parvan17} and Eq.~(8) of Ref.~\cite{Parvan16}. It should be stressed that if we transform the parameter $q\to 1/q$ in Eq.~(\ref{66}), we exactly obtain the transverse momentum distribution of the Tsallis-factorized statistics named the Tsallis distribution~\cite{Cleymans2012} (see Eq.~(56) in Ref.~\cite{Cleymans2012}), which is widely used to fit the experimental data in high energy physics. Thus, the phenomenological Tsallis distribution~\cite{Cleymans2012} for the Maxwell-Boltzmann statistics of particles exactly corresponds to the zeroth term approximation of the Tsallis-1 statistics~\cite{Tsal88} under the transformation $q\to 1/q$. However, the phenomenological Tsallis distributions for the Fermi-Dirac and Bose-Einstein statistics of particles introduced in Ref.~\cite{Cleymans2012} do not correspond to the Tsallis statistics~\cite{Tsal88} (cf. Eqs.~(31) and (33)  of Ref.~\cite{Cleymans2012} along with Eq.~(\ref{56}) of the present paper). See Appendix~\ref{ap1}.

\section{Tsallis unnormalized statistics in the grand canonical ensemble}\label{sec3}

\subsection{General formalism}
The Tsallis unnormalized statistics or the Tsallis-$2$ statistics~\cite{Tsal88} is defined by the generalized entropy, which is the same as the entropy of the Tsallis-$1$ statistics (\ref{1}), with the probabilities $p_{i}$ of the microstates normalized to unity~\cite{Tsal88,Tsal98}
\begin{equation}\label{67}
    S = - \sum\limits_{i} p_{i}^{q} \left(\frac{p_{i}^{1-q}-1}{1-q}\right), \qquad  \sum\limits_{i} p_{i}=1
\end{equation}
and by the generalized expectation values~\cite{Tsal88,Tsal98}
\begin{equation}\label{68}
   \langle A \rangle = \sum\limits_{i} p_{i}^{q} A_{i},
\end{equation}
where $q\in\mathbf{R}$ is a real parameter taking values $0<q<\infty$. The main difference between the Tsallis normalized and Tsallis unnormalized statistics is the definition of their expectation values. In the Tsallis unnormalized statistics, the generalized expectation values (\ref{68}) are not consistent with the norm equation (\ref{67}). Note that in the Gibbs limit $q\to 1$, the entropy (\ref{67}) recovers the Boltzmann-Gibbs-Shannon entropy, $S=-\sum_{i} p_{i} \ln p_{i}$, and the Tsallis unnormalized statistics is reduced to the Boltzmann-Gibbs statistics.

The thermodynamic potential of the grand canonical ensemble of the Tsallis unnormalized statistics can be written as~\cite{Parvan17}
\begin{equation}\label{69}
 \Omega = \langle H \rangle -TS-\mu \langle N \rangle = \sum\limits_{i}  p_{i}^{q} \left[E_{i}-\mu N_{i} + T \frac{p_{i}^{1-q}-1}{1-q}\right],
\end{equation}
where $\langle H \rangle=\sum_{i}  p_{i}^{q} E_{i}$ is the mean energy of the system and $\langle N \rangle=\sum_{i}  p_{i}^{q} N_{i}$ is the mean number of particles of the system.

Using the method of the Lagrange multipliers (\ref{4})--(\ref{6}) and Eq.~(\ref{69}), we get~\cite{Tsal98,Parvan17}
\begin{eqnarray}\label{70}
p_{i} &=& \frac{1}{Z}  \left[1-(1-q) \frac{E_{i}-\mu N_{i}}{T} \right]^{\frac{1}{1-q}}, \\ \label{71}
    Z &=& \sum\limits_{i} \left[1-(1-q) \frac{E_{i}-\mu N_{i}}{T} \right]^{\frac{1}{1-q}},
\end{eqnarray}
where $Z\equiv [(1-(1-q)\lambda/T)/q]^{1/(1-q)}$ is the norm function like the partition function related to the Lagrange multiplier $\lambda$, which is fixed by the norm equation of probabilities given in  Eq.~(\ref{67})~\cite{Tsal98,Parvan17}.

The statistical averages (\ref{68}) for the Tsallis unnormalized statistics in the grand canonical ensemble can be rewritten in the general form as~\cite{Tsal98,ParvanBaldin}
\begin{equation}\label{72}
   \langle A \rangle = \frac{1}{Z^{q}}\sum\limits_{i} A_{i} \left[1-(1-q)\frac{E_{i}-\mu N_{i}}{T}\right]^{\frac{q}{1-q}},
\end{equation}
where $Z$ is calculated from Eq.~(\ref{71}).

Let us rewrite the probabilities of microstates (\ref{70}), the partition function (\ref{71}) and the statistical averages (\ref{72}) in the integral representation. Using Eqs.~(\ref{10}) and (\ref{11}) for $q>1$ and $q<1$, respectively, we obtain the formulas for the integral representation of the probabilities of microstates (\ref{70}) as
\begin{eqnarray}\label{73}
  p_{i}&=&\frac{1}{Z}\frac{1}{\Gamma\left(\frac{1}{q-1}\right)} \int\limits_{0}^{\infty} t^{\frac{1}{q-1}-1} e^{-t\left[1-(1-q)\frac{E_{i}-\mu N_{i}}{T}\right]} dt \nonumber \\ && \mathrm{for} \quad q>1
\end{eqnarray}
and
\begin{eqnarray}\label{74}
  p_{i}&=&\frac{1}{Z}\Gamma\left(\frac{2-q}{1-q}\right)  \frac{i}{2\pi}  \oint\limits_{C} (-t)^{\frac{1}{q-1}-1} \nonumber \\ &\times& e^{-t\left[1-(1-q)\frac{E_{i}-\mu N_{i}}{T}\right]} dt \quad  \mathrm{for} \quad q<1.
\end{eqnarray}
These equations link the probability distribution of the Tsallis unnormalized statistics with the probability distribution of the Boltzmann-Gibbs statistics. The partition function (\ref{71}) in the integral representation can be written as
\begin{eqnarray}\label{75}
 Z&=&\frac{1}{\Gamma\left(\frac{1}{q-1}\right)} \int\limits_{0}^{\infty} t^{\frac{1}{q-1}-1} e^{-t\left[1-(1-q)\frac{\Omega_{G}\left(\beta'\right)}{T}\right]} dt \nonumber \\ && \mathrm{for} \quad q>1
\end{eqnarray}
and
\begin{eqnarray}\label{76}
  Z &=& \Gamma\left(\frac{2-q}{1-q}\right)  \frac{i}{2\pi} \oint\limits_{C} (-t)^{\frac{1}{q-1}-1} e^{-t\left[1-(1-q)\frac{\Omega_{G}\left(\beta'\right)}{T}\right]} dt \nonumber \\ && \mathrm{for} \quad q<1,
\end{eqnarray}
where $\beta'=t(q-1)/T$ and $\Omega_{G}\left(\beta'\right)$ is the thermodynamic potential of the Boltzmann-Gibbs statistics defined in Eq.~(\ref{16}). Equations~(\ref{75}) and (\ref{76}) link the partition function $Z$ of the Tsallis unnormalized statistics with the thermodynamic potential of the Boltzmann-Gibbs statistics.

Let us rewrite the statistical averages (\ref{72}) in the integral representation. Using Eqs.~(\ref{10}) and (\ref{11}) for $q>1$ and $q<1$, respectively, we obtain
\begin{eqnarray}\label{77}
  \langle A \rangle &=& \frac{1}{Z^{q}}\frac{1}{\Gamma\left(\frac{q}{q-1}\right)} \int\limits_{0}^{\infty} t^{\frac{1}{q-1}} e^{-t\left[1-(1-q)\frac{\Omega_{G}\left(\beta'\right)}{T}\right]} \nonumber \\ &&  \langle A \rangle_{G}\left(\beta'\right) dt \quad  \mathrm{for} \quad q>1
\end{eqnarray}
and
\begin{eqnarray}\label{78}
   \langle A \rangle &=& \frac{1}{Z^{q}}\Gamma\left(\frac{1}{1-q}\right)  \frac{i}{2\pi} \oint\limits_{C} dt (-t)^{\frac{1}{q-1}} \nonumber \\ && e^{-t\left[1-(1-q)\frac{\Omega_{G}\left(\beta'\right)}{T}\right]} \langle A \rangle_{G}\left(\beta'\right)   \quad \mathrm{for} \quad q<1, \;\;\;\;
\end{eqnarray}
where $\langle A \rangle_{G}\left(\beta'\right)$ is the statistical averages of the Boltzmann - Gibbs statistics defined in Eq.~(\ref{19}). Equations~(\ref{77}) and (\ref{78}) link the statistical averages of the Tsallis unnormalized statistics with the corresponding statistical averages of the Boltzmann-Gibbs statistics.

\subsection{Transverse momentum distribution in the Tsallis-2 statistics}
\subsubsection{Exact results}
\paragraph{General case}
The transverse momentum distribution in the Tsallis unnormalized statistics in the grand canonical ensemble is derived from Eqs.~(\ref{77}) and (\ref{78}). For the ideal gas with the Fermi-Dirac, Bose-Einstein and Maxwell-Boltzmann statistics of particles, it can be written in the form of the series expansion as (see Appendix~\ref{ap2})
\begin{eqnarray}\label{91}
  \frac{d^{2}N}{dp_{T}dy} &=& \frac{gV}{(2\pi)^{2}} p_{T}  m_{T} \cosh y  \sum\limits_{n=0}^{\infty}  \frac{1}{n!Z^{q}\Gamma\left(\frac{q}{q-1}\right)} \nonumber \\ && \int\limits_{0}^{\infty} t^{\frac{1}{q-1}} e^{-t}    \frac{\left(-\beta'\Omega_{G}\left(\beta'\right)\right)^{n}}{e^{\beta' (m_{T} \cosh y-\mu)}+\eta} dt \nonumber \\ &&  \mathrm{for} \quad q>1
\end{eqnarray}
and
\begin{eqnarray}\label{92}
 \frac{d^{2}N}{dp_{T}dy} &=& \frac{gV}{(2\pi)^{2}} p_{T}  m_{T} \cosh y \sum\limits_{n=0}^{\infty}  \frac{\Gamma\left(\frac{1}{1-q}\right)}{n!Z^{q}}    \frac{i}{2\pi} \nonumber \\ && \oint\limits_{C} (-t)^{\frac{1}{q-1}} e^{-t} \frac{\left(-\beta'\Omega_{G}\left(\beta'\right)\right)^{n}}{e^{\beta' (m_{T} \cosh y-\mu)}+\eta} dt \nonumber \\ &&  \mathrm{for} \quad q<1.
\end{eqnarray}
The partition function (\ref{71}) in the integral representation (\ref{75}) and (\ref{76}) can be rewritten in the form of the series expansion as (see Appendix~\ref{ap2})
\begin{eqnarray}\label{80}
 Z &=& \sum\limits_{n=0}^{\infty}  \frac{1}{n!\Gamma\left(\frac{1}{q-1}\right)} \int\limits_{0}^{\infty} t^{\frac{1}{q-1}-1} e^{-t} \left(-\beta'\Omega_{G}\left(\beta'\right)\right)^{n} dt  \nonumber \\ &&
 \mathrm{for} \quad q>1
\end{eqnarray}
and
\begin{eqnarray}\label{81}
 Z &=& \sum\limits_{n=0}^{\infty} \frac{\Gamma\left(\frac{2-q}{1-q}\right)}{n!}    \frac{i}{2\pi} \oint\limits_{C} (-t)^{\frac{1}{q-1}-1} e^{-t} \left(-\beta'\Omega_{G}\left(\beta'\right)\right)^{n} dt
 \nonumber \\ && \mathrm{for} \quad q<1,
\end{eqnarray}
where $\Omega_{G}\left(\beta'\right)$ is defined in Eq.~(\ref{23}).

\begin{figure*}[!htb]
\vspace*{+1cm}
\minipage{0.42\textwidth}
\includegraphics[width=\linewidth]{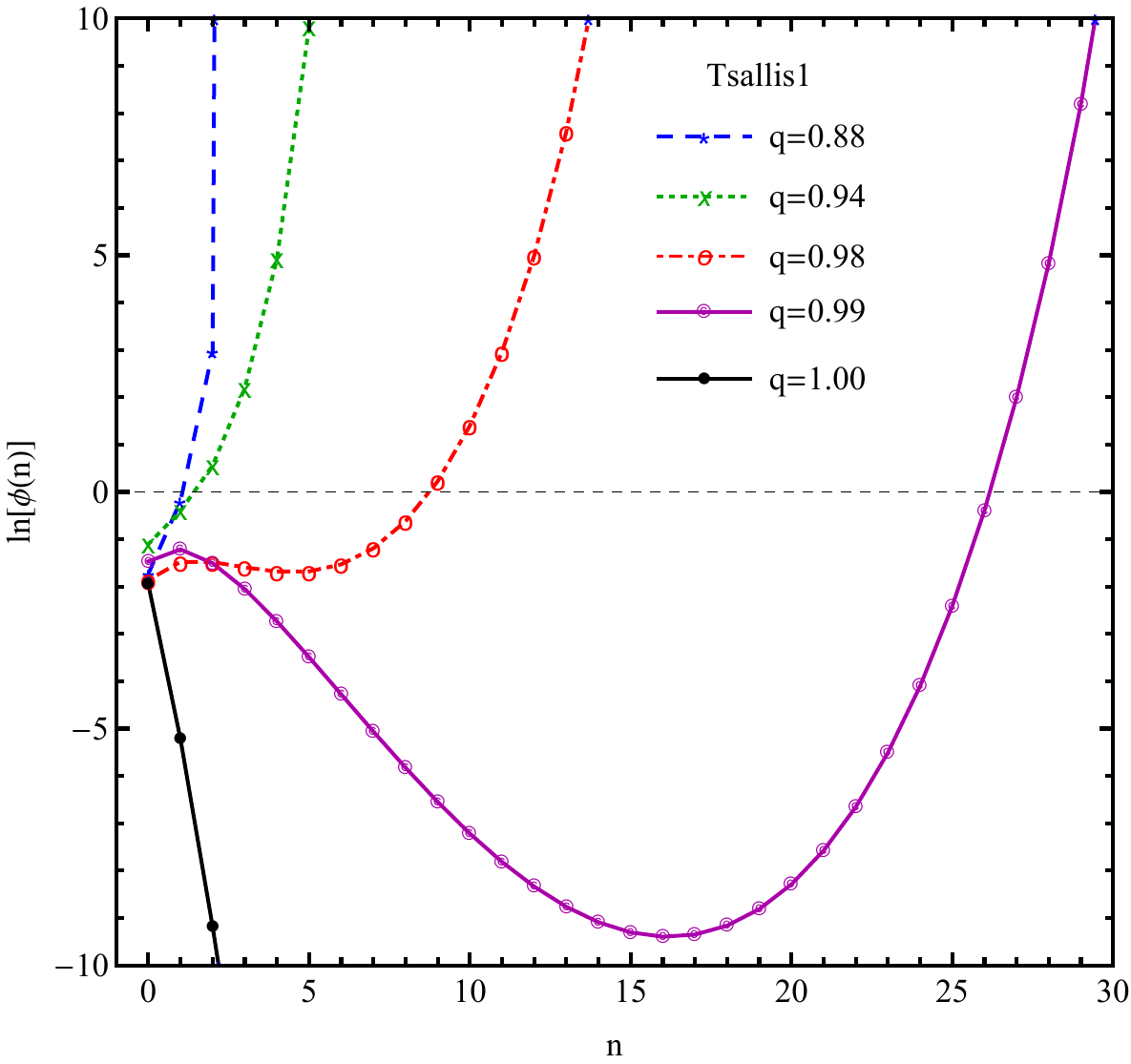}
\endminipage\hfill
\minipage{0.42\textwidth}
\vspace*{0.0cm}
\hspace*{0cm}
\includegraphics[width=\linewidth]{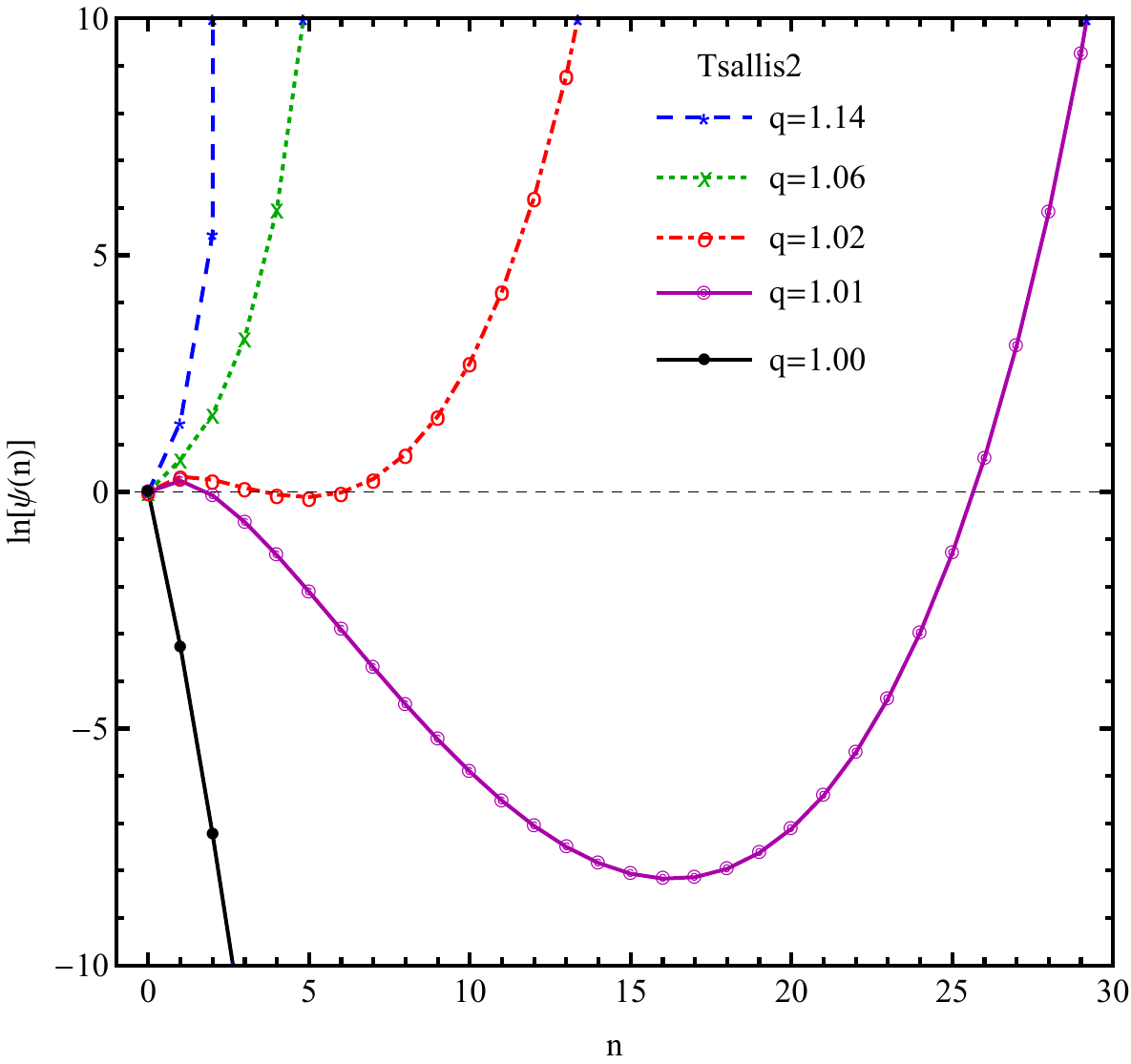}
\endminipage\hfill
\caption{(Color online) The functions $\ln\phi(n)$ and $\ln\psi(n)$ vs $n$ (number of terms) in the Tsallis-1 and Tsallis-2 statistics for the Maxwell-Boltzmann statistics of particles. Temperature $T=82$ MeV, chemical potential $\mu=0$, radius $R=4$ fm and mass $m=139.57$ MeV (pion mass). $\phi(n)$ and $\psi(n)$ are given by Eqs.~(\ref{39}) and (\ref{93}), respectively.}
\label{fig1}
\end{figure*}

\paragraph{Maxwell-Boltzmann statistics of particles}
Taking the \\ limit $\eta\to 0$ the formulas for the Maxwell-Boltzmann statistics of particles can be rewritten explicitly. Substituting Eq.~(\ref{38}) into Eqs.~(\ref{80}) and (\ref{81}), we obtain the partition function for the Maxwell-Boltzmann statistics of particles in the Tsallis unnormalized statistics as
\begin{eqnarray}\label{93}
 Z &=& \sum\limits_{n=0}^{\infty} \frac{\omega^{n}}{n!} \frac{1}{\Gamma\left(\frac{1}{q-1}\right)} \nonumber \\ &\times& \int\limits_{0}^{\infty} t^{\frac{1}{q-1}-1-n} e^{-t\left[1+(1-q)\frac{\mu n}{T}\right]} \nonumber \\ &\times& \left(K_{2}\left(\frac{t(q-1)m}{T} \right)\right)^{n} dt  \quad \mathrm{for} \quad q>1 \nonumber
\end{eqnarray}
which can be briefly written as
\begin{equation}
 Z = \sum\limits_{n=0}^{\infty} \psi(n)
\end{equation}
and
\begin{eqnarray}\label{94}
Z &=& \sum\limits_{n=0}^{\infty} \frac{(-\omega)^{n}}{n!}  \Gamma\left(\frac{2-q}{1-q}\right)  \frac{i}{2\pi} \nonumber \\ &\times& \oint\limits_{C} (-t)^{\frac{1}{q-1}-1-n} e^{-t\left[1+(1-q)\frac{\mu n}{T}\right]} \nonumber \\ &\times& \left(K_{2}\left(\frac{t(q-1)m}{T} \right)\right)^{n} dt \quad \mathrm{for} \quad q<1,
\end{eqnarray}
where
\begin{equation}\label{95}
  \omega=\frac{gVTm^{2}}{2\pi^{2}} \frac{1}{q-1}.
\end{equation}
Note that in the ultrarelativistic limit $(m=0)$, Eqs.~(\ref{93}) and (\ref{94}) recover Eq.~(63) of Ref.~\cite{Parvan17}.

Substituting Eq.~(\ref{38}) into Eqs.~(\ref{91}) and (\ref{92}) and taking $\eta=0$, we obtain the transverse momentum distribution for the Maxwell-Boltzmann statistics of particles in the Tsallis unnormalized statistics as
\begin{eqnarray}\label{102}
  \frac{d^{2}N}{dp_{T}dy} &=& \frac{gV}{(2\pi)^{2}} p_{T}  m_{T} \cosh y  \sum\limits_{n=0}^{\infty} \frac{\omega^{n}}{n!}\frac{1}{Z^{q}} \frac{1}{\Gamma\left(\frac{q}{q-1}\right)} \nonumber \\
  &\times& \int\limits_{0}^{\infty} t^{\frac{1}{q-1}-n}
 e^{-t\left[1-(1-q)\frac{m_{T} \cosh y-\mu(n+1)}{T}\right]} \nonumber \\ &\times& \left(K_{2}\left(\frac{t(q-1)m}{T} \right)\right)^{n} dt \quad \mathrm{for} \quad q>1
\end{eqnarray}
and
\begin{eqnarray}\label{103}
 \frac{d^{2}N}{dp_{T}dy} &=& \frac{gV}{(2\pi)^{2}} p_{T}  m_{T} \cosh y \sum\limits_{n=0}^{\infty} \frac{(-\omega)^{n}}{n!} \frac{1}{Z^{q}} \Gamma\left(\frac{1}{1-q}\right) \nonumber \\
   &\times& \frac{i}{2\pi} \oint\limits_{C} (-t)^{\frac{1}{q-1}-n} e^{-t\left[1-(1-q)\frac{m_{T} \cosh y-\mu(n+1)}{T}\right]} \nonumber \\ &\times& \left(K_{2}\left(\frac{t(q-1)m}{T} \right)\right)^{n} dt \quad \mathrm{for} \quad q<1.
\end{eqnarray}
In the ultrarelativistic limit $(m=0)$, Eqs.~(\ref{102}) and (\ref{103}) recover Eq.~(73) of Ref.~\cite{Parvan17}. See Appendix~\ref{ap2}.

\begin{figure*}[!htb]
\vspace*{+1cm}
\minipage{0.42\textwidth}
\includegraphics[width=\linewidth]{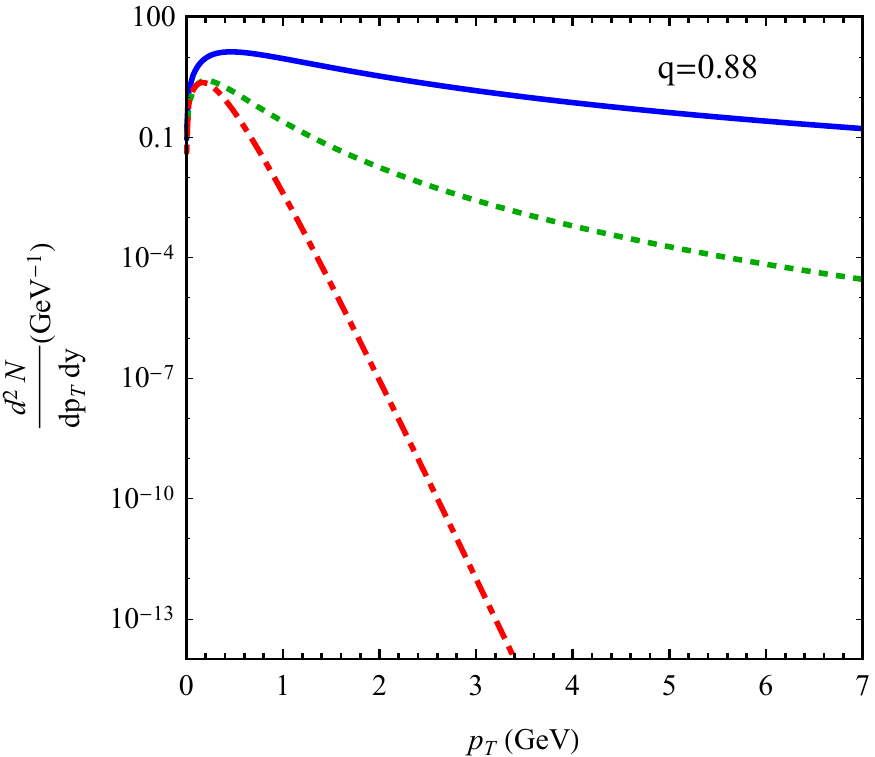}
\endminipage\hfill
\minipage{0.42\textwidth}
\vspace*{-0cm}
\hspace*{0cm}
\includegraphics[width=\linewidth]{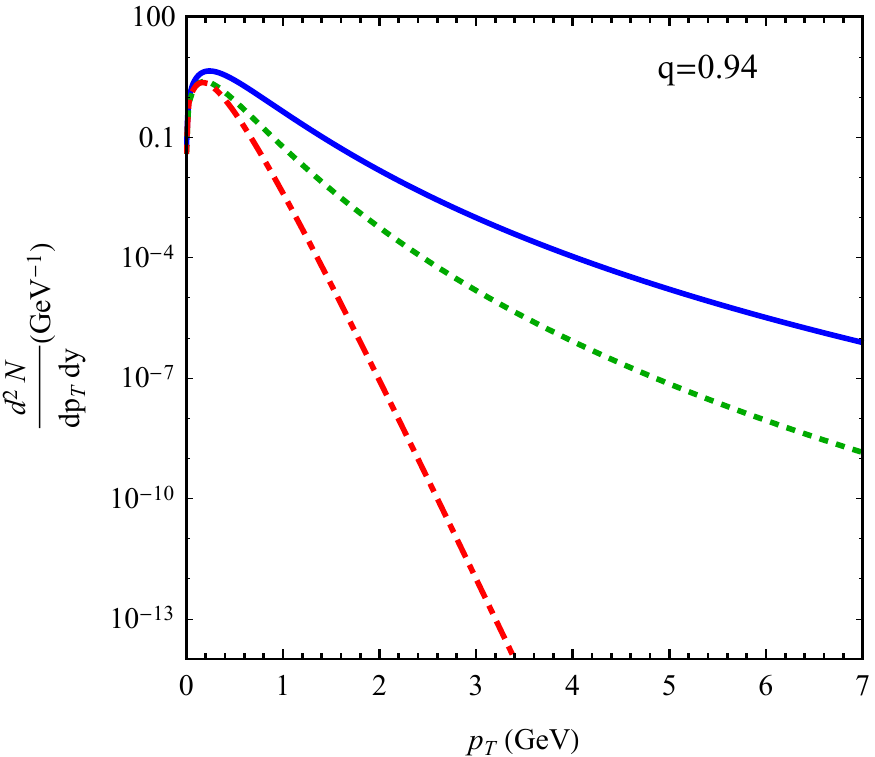}
\endminipage\hfill
\minipage{0.42\textwidth}
\vspace*{+0cm}
\hspace*{-0.1cm}
\includegraphics[width=\linewidth]{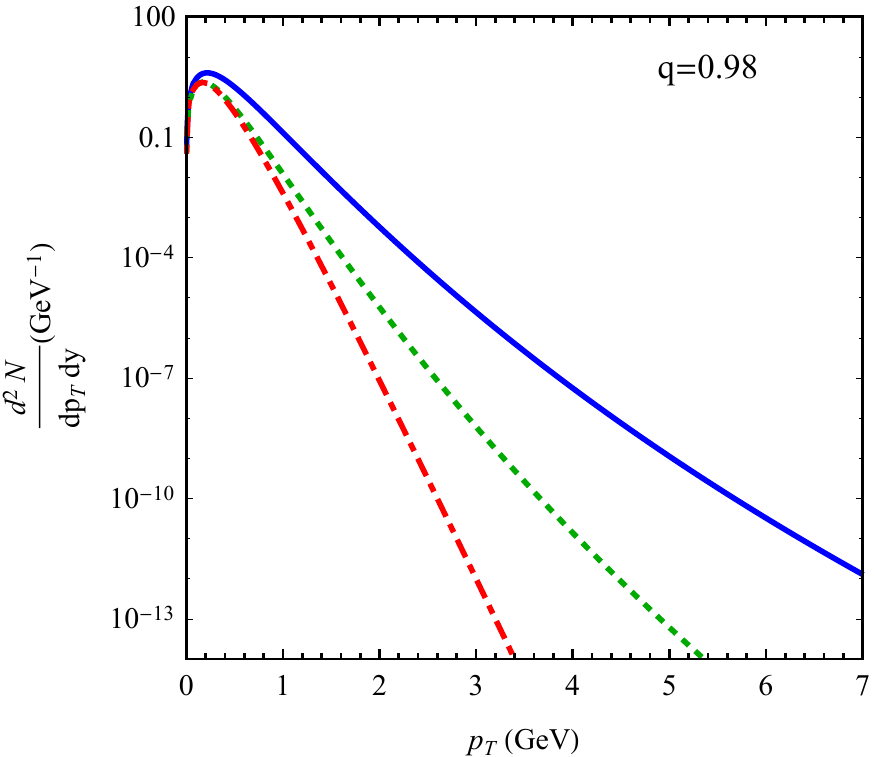}
\endminipage\hfill
\minipage{0.42\textwidth}
\vspace*{+0cm}
\hspace*{0cm}
\includegraphics[width=\linewidth]{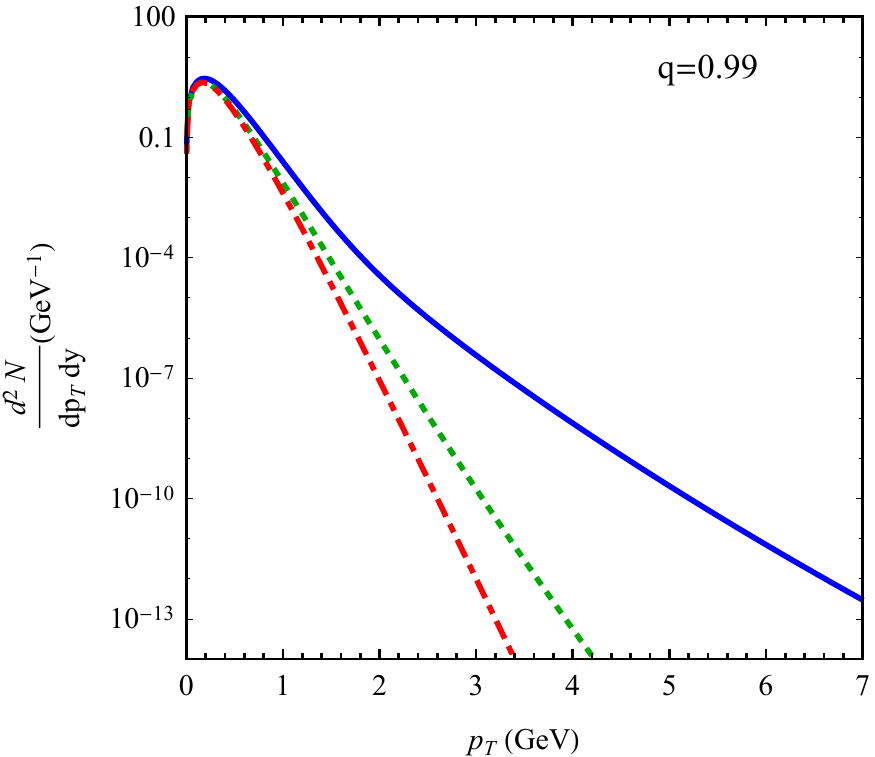}
\endminipage\hfill
\caption{(Color online) The spectra of the Maxwell-Boltzmann massive particles in the Tsallis-1 statistics at mid-rapidity ($y=0$) for different values of the entropic parameter $q$. Temperature $T=82$ MeV, chemical potential $\mu=0$, radius $R=4$ fm and mass $m=139.57$ MeV (pion mass). The solid, doted and dot-dashed lines correspond to the exact Tsallis-1 statistics, zeroth-term approximation and the Boltzmann-Gibbs statistics $(q=1)$, respectively.}
\label{fig2}
\end{figure*}

\subsubsection{Zeroth term approximation}
Taking $n=0$ in Eqs.~(\ref{80}) and (\ref{81}) and using Eqs.~(\ref{10}) and (\ref{11}), we obtain that the partition function $Z=1$. Considering only the zeroth term in Eqs.~(\ref{91}) and (\ref{92}) with $Z=1$ and using Eqs.~(\ref{10}), (\ref{11}) and (\ref{54}), we obtain
\begin{eqnarray}\label{116}
\frac{d^{2}N}{dp_{T}dy} &=& \frac{gV}{(2\pi)^{2}} p_{T}  m_{T} \cosh y \nonumber \\ &\times&  \sum\limits_{k=0}^{\infty} (-\eta)^{k}   \left[1+(k+1) (q-1) \frac{m_{T} \cosh y-\mu}{T} \right]^{\frac{q}{1-q}} \nonumber \\ &&   \mathrm{for} \quad \eta=-1,0,1
\end{eqnarray}
or
\begin{eqnarray}\label{117}
\frac{d^{2}N}{dp_{T}dy} &=& \frac{gV}{(2\pi)^{2}} p_{T}  m_{T} \cosh y \nonumber \\ &\times& \left[1+ (q-1) \frac{m_{T} \cosh y-\mu}{T} \right]^{\frac{q}{1-q}} \nonumber \\ &&   \mathrm{for} \quad \eta=0.
\end{eqnarray}
In the ultrarelativistic limit $(m=0)$, Eq.~(\ref{117}) recovers Eq.~(80) of Ref.~\cite{Parvan17} and Eq.~(15) from Ref.~\cite{ParvanBaldin}.
It should be stressed that Eq.~(\ref{117}) exactly coincides with the transverse momentum distribution of the Maxwell-Boltzmann statistics of particles of the Tsallis-factorized statistics defined in Ref.~\cite{Cleymans2012} (see Eq.~(56) in Ref.~\cite{Cleymans2012}). Thus, we have proved that in the case of the Maxwell-Boltzmann statistics of particles, the phenomenological Tsallis distribution introduced in Ref.~\cite{Cleymans2012} is the transverse momentum distribution of the Tsallis unnormalized statistics in the zeroth term approximation. However, the Tsallis distributions for the Fermi-Dirac and Bose-Einstein statistics, which were also introduced in Ref.~\cite{Cleymans2012} by the ansatz, do not correspond to the Tsallis unnormalized statistics (cf. Eqs.~(31) and (33) from ref.~\cite{Cleymans2012} with Eq.~(\ref{107}) of the present paper). See Appendix~\ref{ap2}. Note that there is a limiting value of $q$ at which the zeroth term approximation and the Tsallis statistics on the whole become divergent~\cite{BhattaCleMog,Parvan17a}.

\begin{figure*}[!htb]
\vspace*{+1cm}
\minipage{0.42\textwidth}
\includegraphics[width=\linewidth]{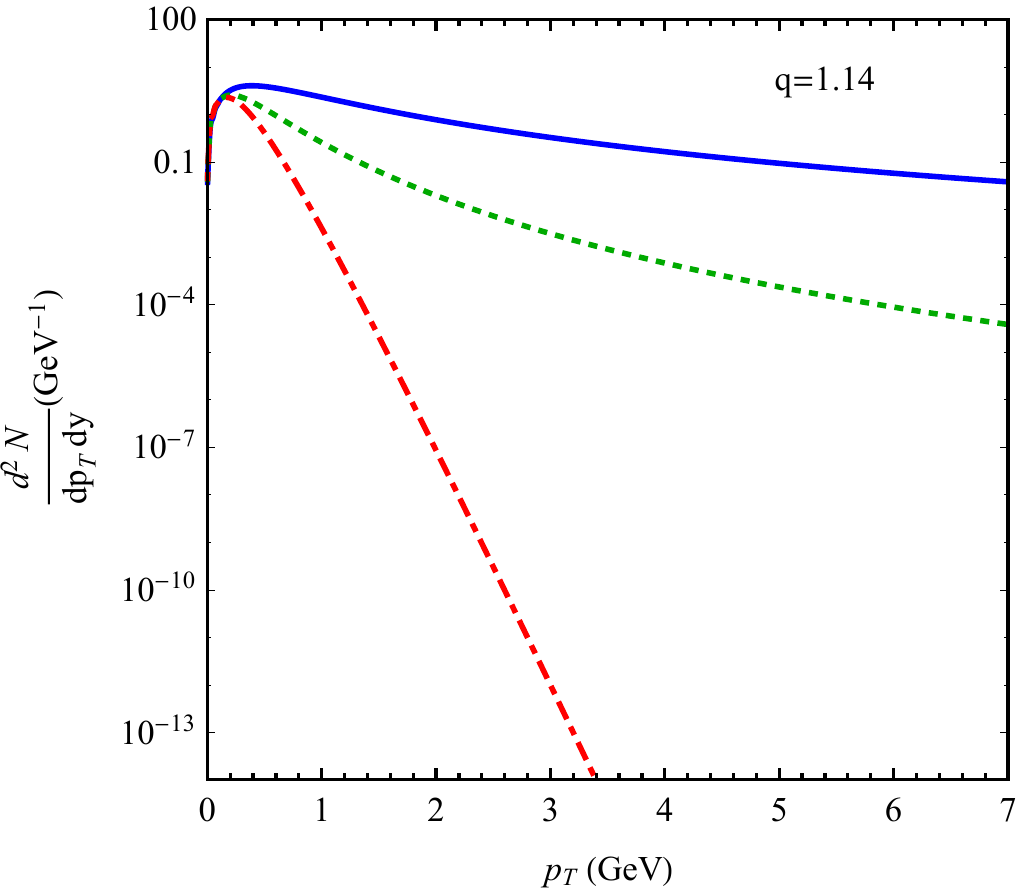}
\endminipage\hfill
\minipage{0.42\textwidth}
\vspace*{-0cm}
\hspace*{-0cm}
\includegraphics[width=\linewidth]{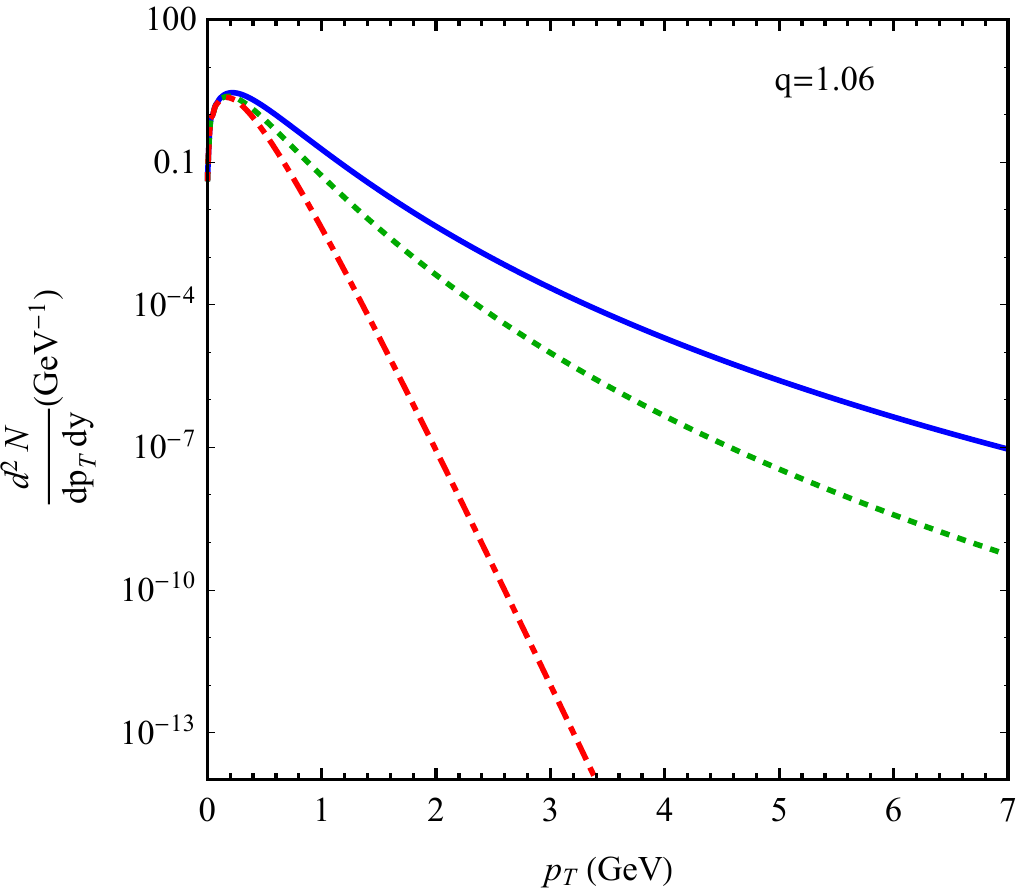}
\endminipage\hfill
\minipage{0.42\textwidth}
\vspace*{+0cm}
\hspace*{-0cm}
\includegraphics[width=\linewidth]{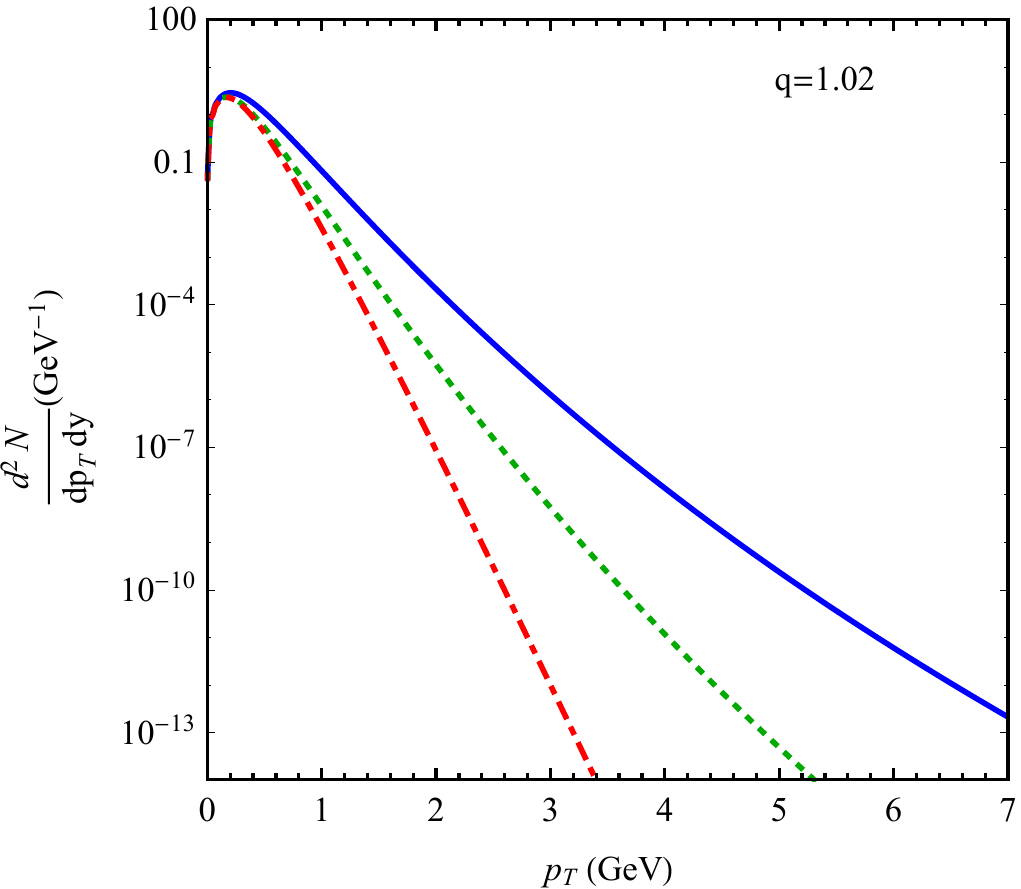}
\endminipage\hfill
\minipage{0.42\textwidth}
\vspace*{+0cm}
\hspace*{-0cm}
\includegraphics[width=\linewidth]{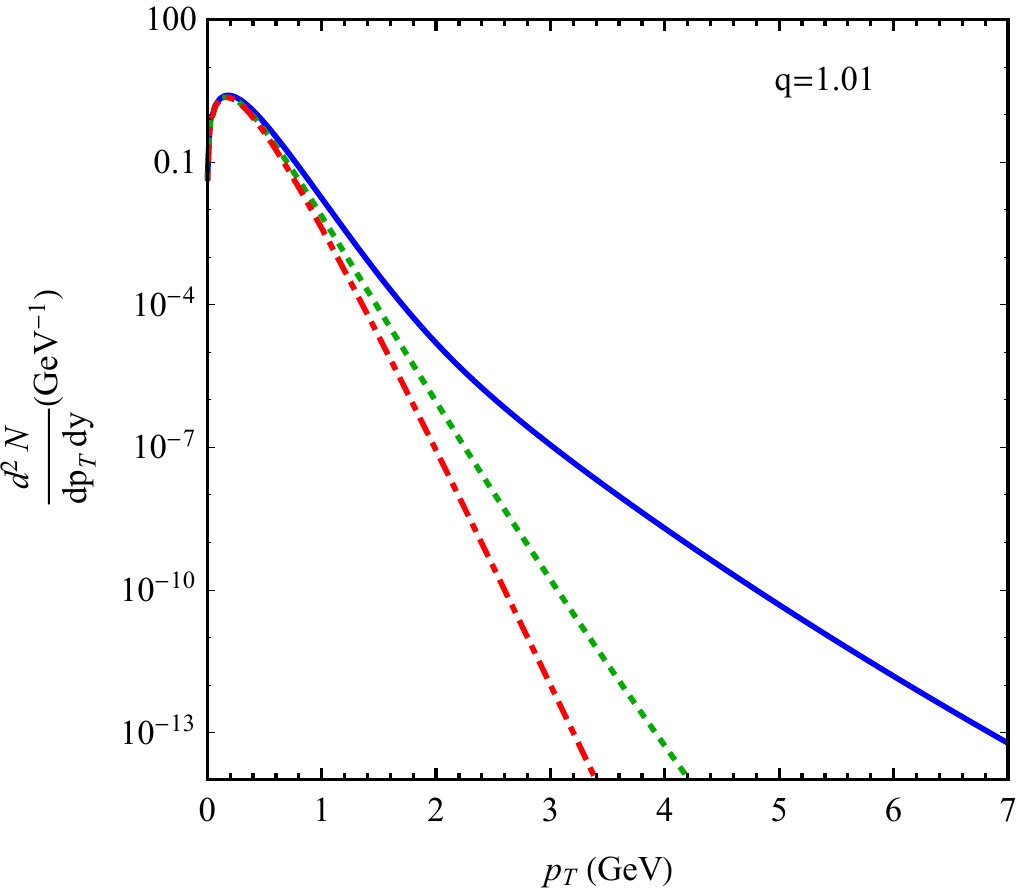}
\endminipage\hfill
\caption{(Color online) The spectra of the Maxwell-Boltzmann massive particles in the Tsallis-2 statistics at mid-rapidity ($y=0$) for different values of the entropic parameter $q$. Temperature $T=82$ MeV, chemical potential $\mu=0$, radius $R=4$ fm and mass $m=139.57$ MeV (pion mass). The solid, dotted and dot-dashed lines correspond to the exact Tsallis-2 statistics, zeroth-term approximation and the Boltzmann-Gibbs statistics $(q=1)$, respectively.}
\label{fig3}
\end{figure*}

\begin{figure}[!htb]
\begin{center}
\includegraphics[width=0.42\textwidth]{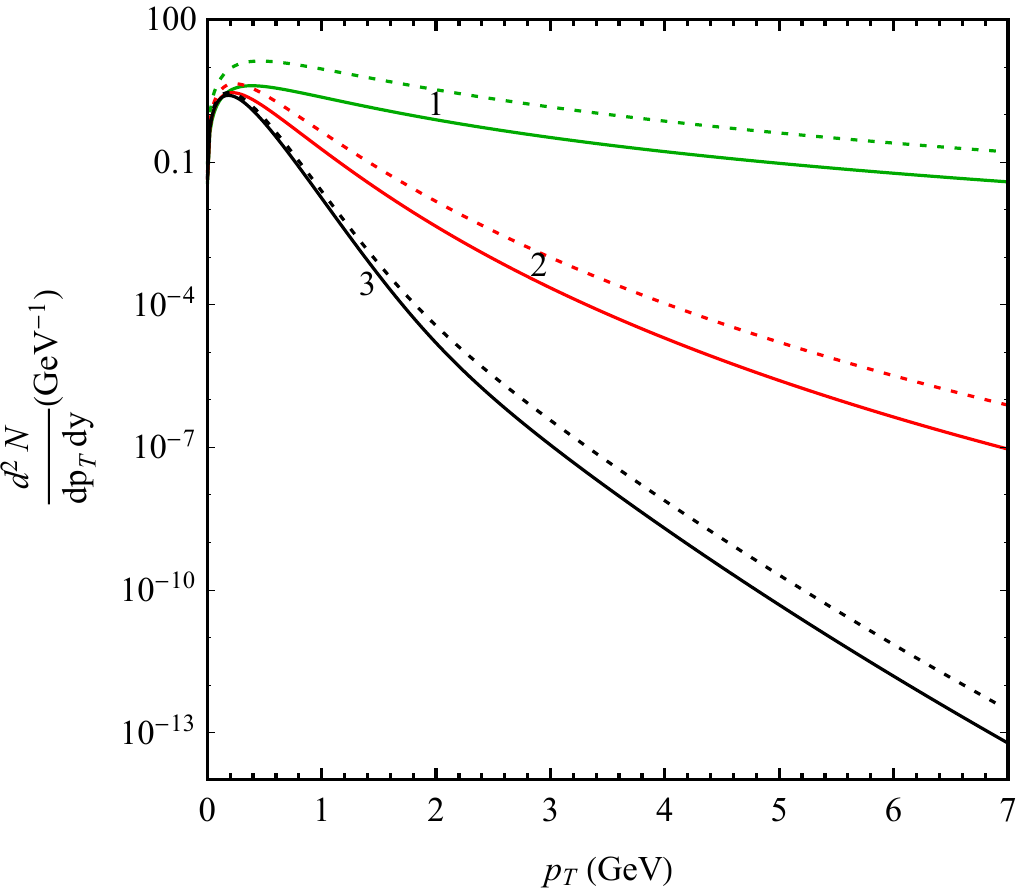}
\caption{(Color online) The spectra of the Maxwell-Boltzmann massive particles in the Tsallis-1 (dotted lines) and Tsallis-2 (solid lines) statistics at mid-rapidity ($y=0$) for different values of the entropic parameter $q$. Temperature $T=82$ MeV, chemical potential $\mu=0$, radius $R=4$ fm and mass $m=139.57$ MeV (pion mass). Lines $1,2$ and $3$ correspond to $q=0.88(1/0.88),0.94 (1/0.94)$ and $0.99 (1/0.99)$ for the exact Tsallis-1 (Tsallis-2) statistics.}
\label{fig4}
\end{center}
\end{figure}

\section{Analysis and results}\label{sec4}
After discussing the general formalisms of the Tsallis-1 and Tsallis-2 statistics, we now proceed to compare the particle spectrum $d^2N/dp_{\mathrm{T}} dy$, which is an experimentally measurable quantity, calculated in these two approaches. We have already noticed that the mathematical expressions for the particle spectra involve infinite summations (see Eqs.~(\ref{36}) and (\ref{91})) which diverge for an arbitrarily large number of terms. Hence, we introduce the procedure of regularizing the particle spectra. For the Tsallis-1 statistics, the upper cut-off (say $n=n_0$, where $n$ is the ordinality of the terms) is found from the minimum of the natural logarithm of $\phi(n)$, which appears while normalizing the probabilities (see Eq.~(\ref{39})). For the Tsallis-2 statistics, $n_0$ is found from the local minimum of the natural logarithm of the function $\psi(n)$, which appears in the description of the partition function (see Eq.~(\ref{93})). For the visual demonstration of the appearance of the minima, we have plotted the variation of $\ln\phi(n)$ and $\ln\psi(n)$ with $n$ for different values of the Tsallis parameter $q$ in Fig.~\ref{fig1}. In the Gibbs limit $q \rightarrow 1$, the functions $\ln\phi(n)$ and $\ln\psi(n)$ are monotonically decreasing with $n$. With decreasing/increasing values of $q$ (depending on whether it is the Tsallis-1 or the Tsallis-2 formalism), prominent local minima appear and after certain values of $n$ the functions start diverging. The prominence of the local minima disappears with decreasing (increasing) $q$ values in the Tsallis-1(Tsallis-2) statistics.

\begin{figure*}[!htb]
\vspace*{+1cm}
\minipage{0.42\textwidth}
\includegraphics[width=\linewidth]{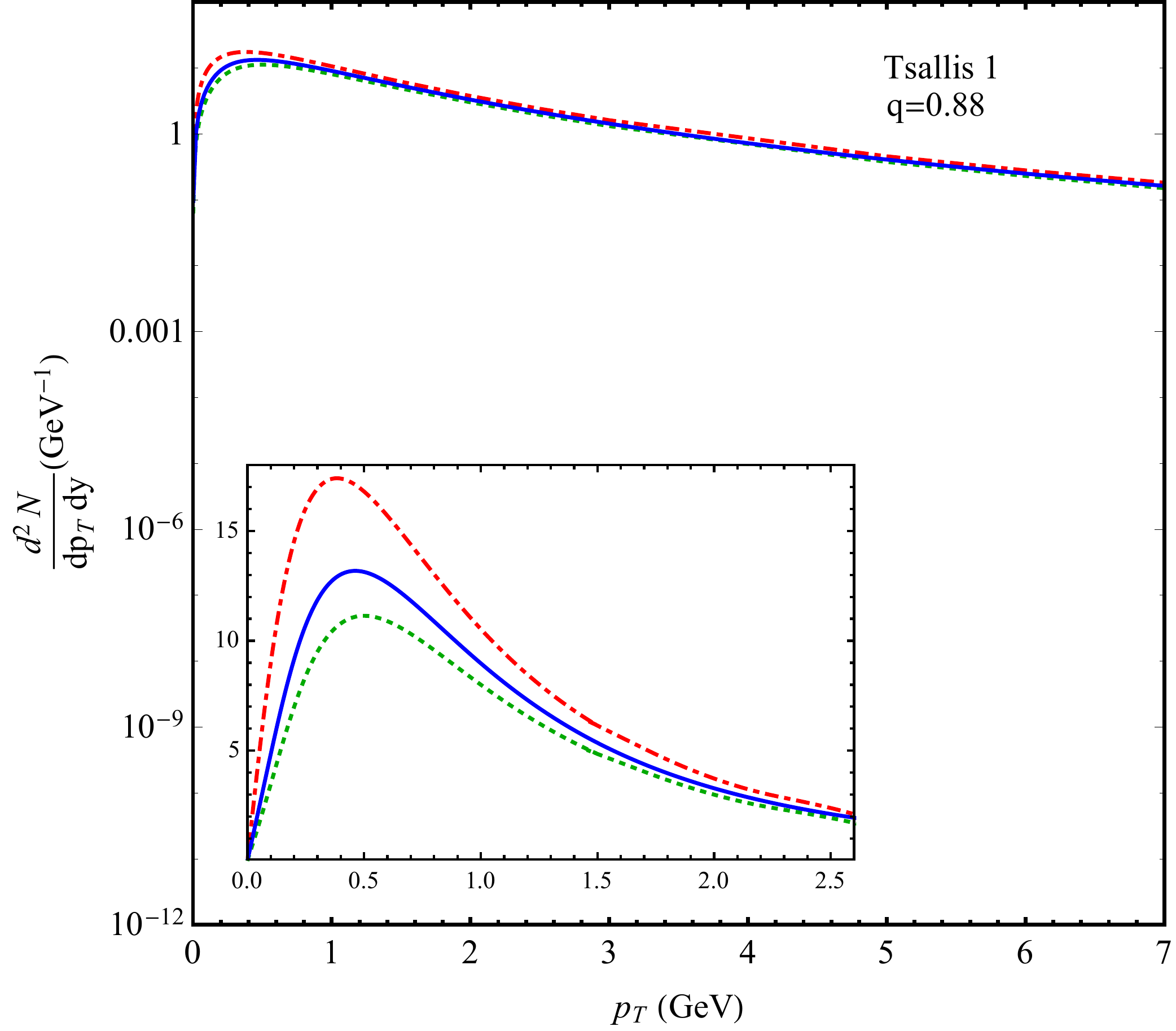}
\endminipage\hfill
\minipage{0.42\textwidth}
\vspace*{-0cm}
\hspace*{-0cm}
\includegraphics[width=\linewidth]{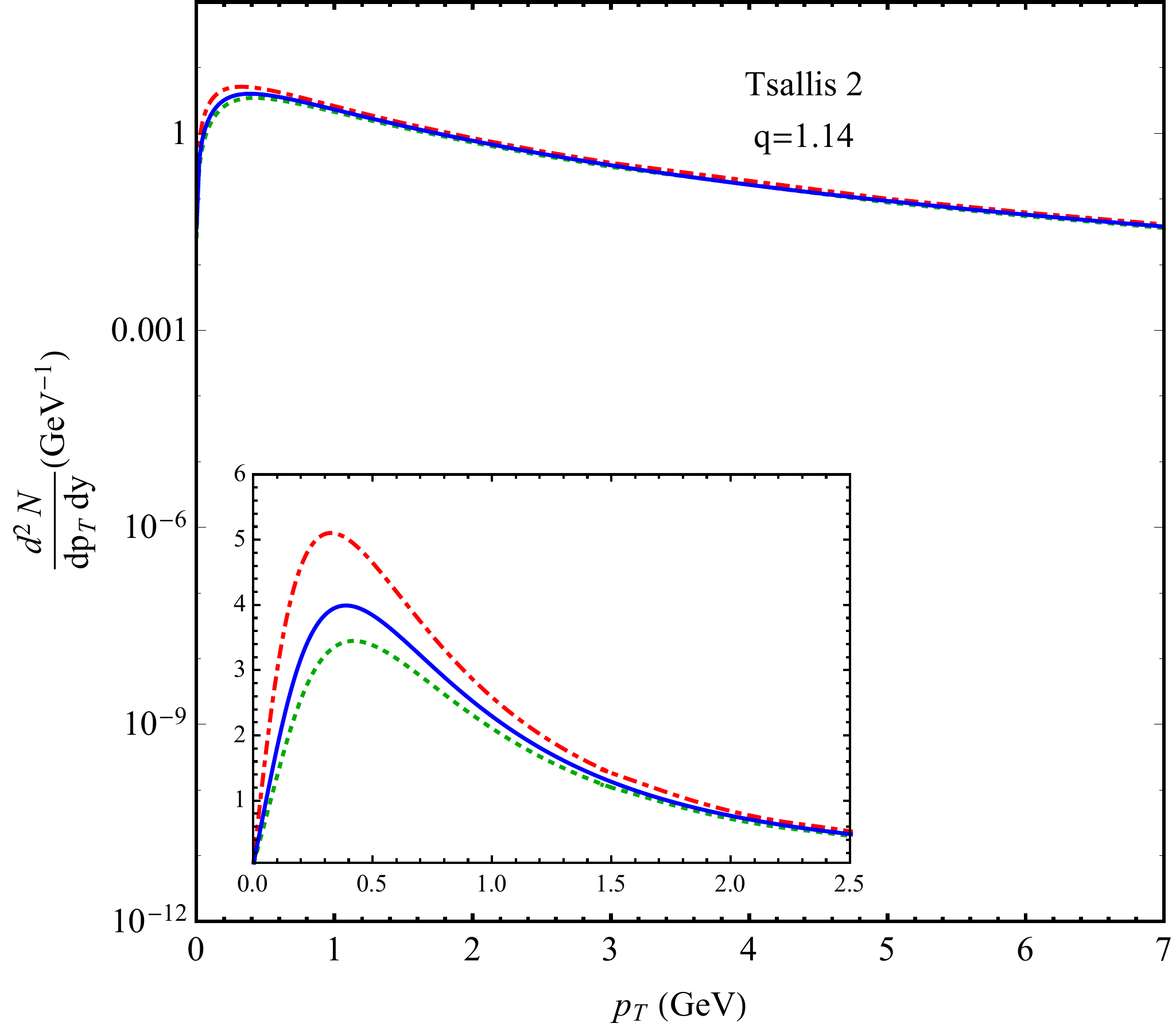}
\endminipage\hfill
\minipage{0.42\textwidth}
\vspace*{+0cm}
\hspace*{-0cm}
\includegraphics[width=\linewidth]{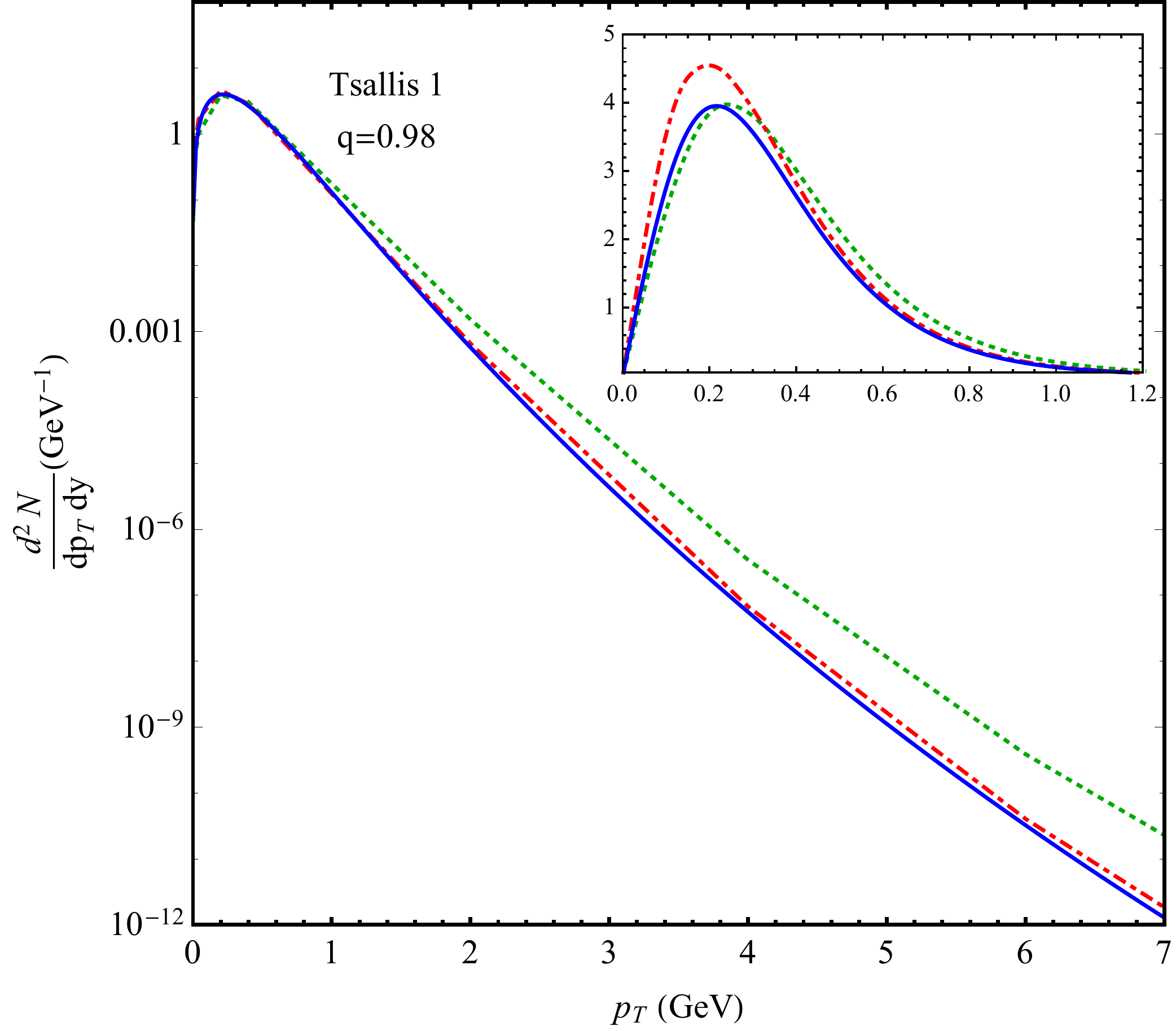}
\endminipage\hfill
\minipage{0.42\textwidth}
\vspace*{+0cm}
\hspace*{-0cm}
\includegraphics[width=\linewidth]{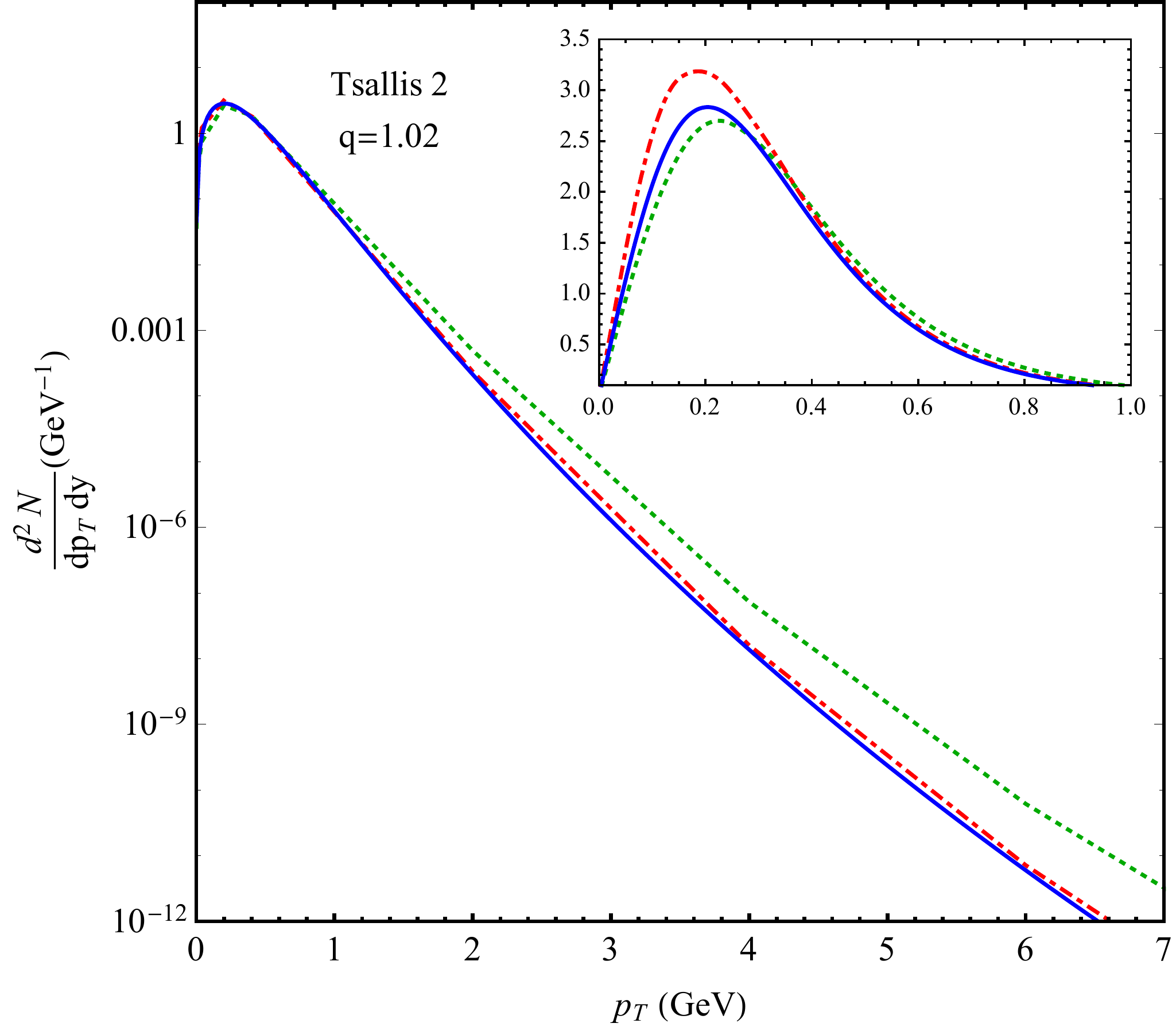}
\endminipage\hfill
\caption{(Color online) Comparison among the Maxwell-Boltzmann (blue solid line) , Fermi-Dirac (green dotted line) and Bose-Einstein (red dot-dashed line) spectra in the Tsallis-1 (left panels) and Tsallis-2 (right panels) exact formalisms at mid-rapidity $(y=0)$ for different values of the entropic parameter $q$. Temperature $T=82$ MeV, chemical potential $\mu=0$ MeV, radius $R=4$ fm and mass $m=139.57$ MeV for all the figures. Inset: the low transverse momentum variation of the spectra.}
\label{fig5}
\end{figure*}

After this preliminary discussion on the regularization scheme of the Tsallis-1 and the Tsallis-2 statistics, we turn our attention to the particle spectra in the Tsallis statistics and the Tsallis-factorized statistics. In Fig.~\ref{fig2} we show the mid-rapidity ($y=0$) spectra (Tsallis Maxwell-Boltzmann as well as Tsallis Maxwell-Boltzmann factorized which is the same as the $n=0$ term in the expansion) of the $\pi^-$ particles for the fixed values of temperature $T=82$ MeV, radius $R=4$ fm and chemical potential $\mu=0$ for four different $q$ values in the Tsallis-1 statistics. It is noticed that the spectra get harder with decreasing values of $q$. This behaviour is not unexpected because smaller $q$ values correspond to the spectra in the LHC region where we get higher momentum particles~\cite{Parvan16}. As opposed to that, larger values of $q$ correspond to the region probed by the experiments like NA61/SHINE, for example, where the high-$p_{T}$ particle production is less~\cite{Parvan16,Parvan17a}. Thus, we have obtained that the Tsallis-1 statistics essentially increases the production of high-$p_{T}$ hadrons in comparison with the phenomenological Tsallis distribution at the same value of $q$.

The same behaviour is noticed in Fig.~\ref{fig3} for the Tsallis-2 statistics so that the larger the $q$ value is, the harder the spectra are. This observation reveals the inverse relationship between the $q$ parameters in the two approaches so that higher $q$ values in Fig.~\ref{fig3} correspond to lower $q$ values in Fig.~\ref{fig2}.

\begin{figure*}[!htb]
\vspace*{+1cm}
\minipage{0.42\textwidth}
\includegraphics[width=\linewidth]{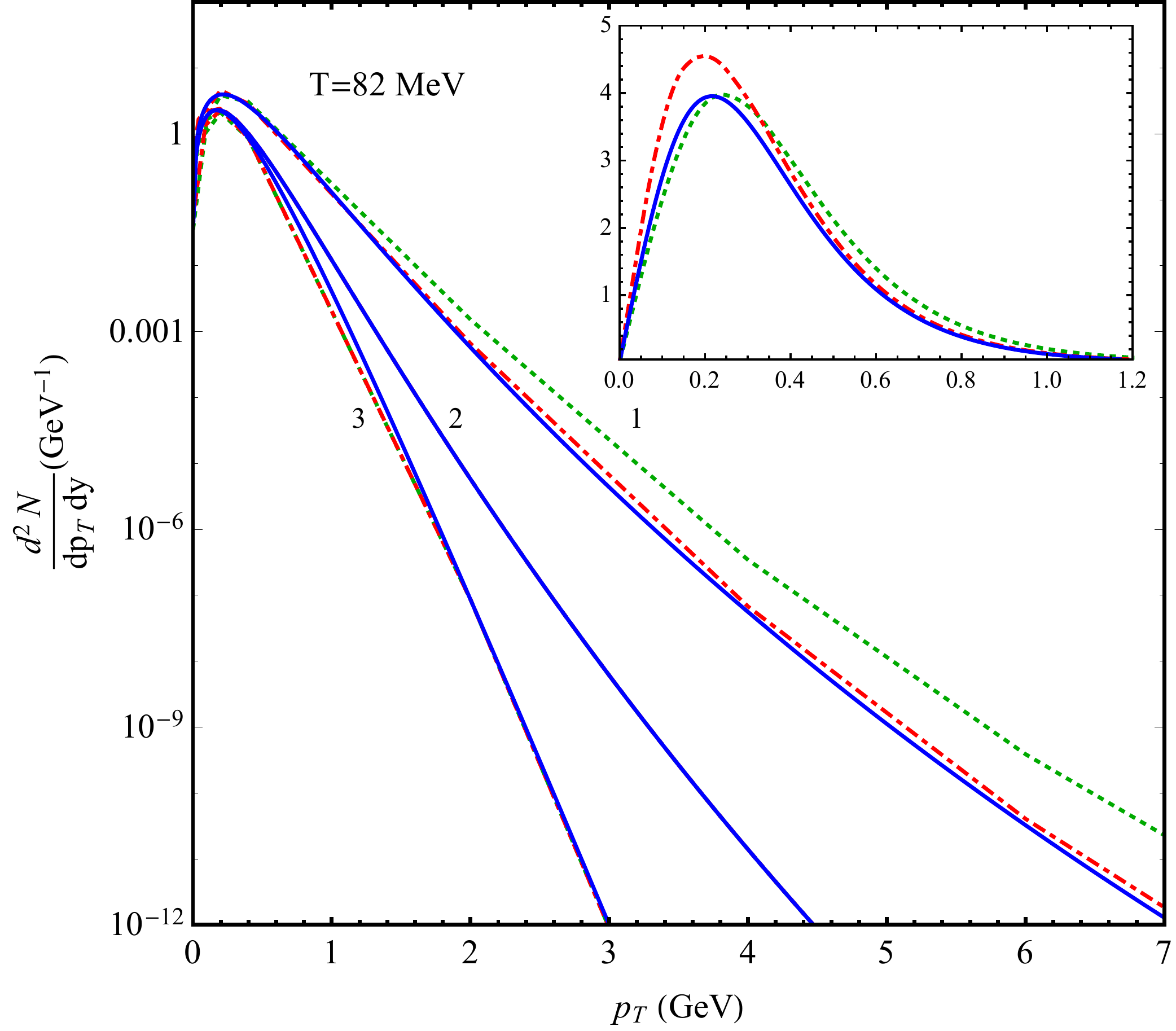}
\endminipage\hfill
\minipage{0.42\textwidth}
\vspace*{-0cm}
\hspace*{-0cm}
\includegraphics[width=\linewidth]{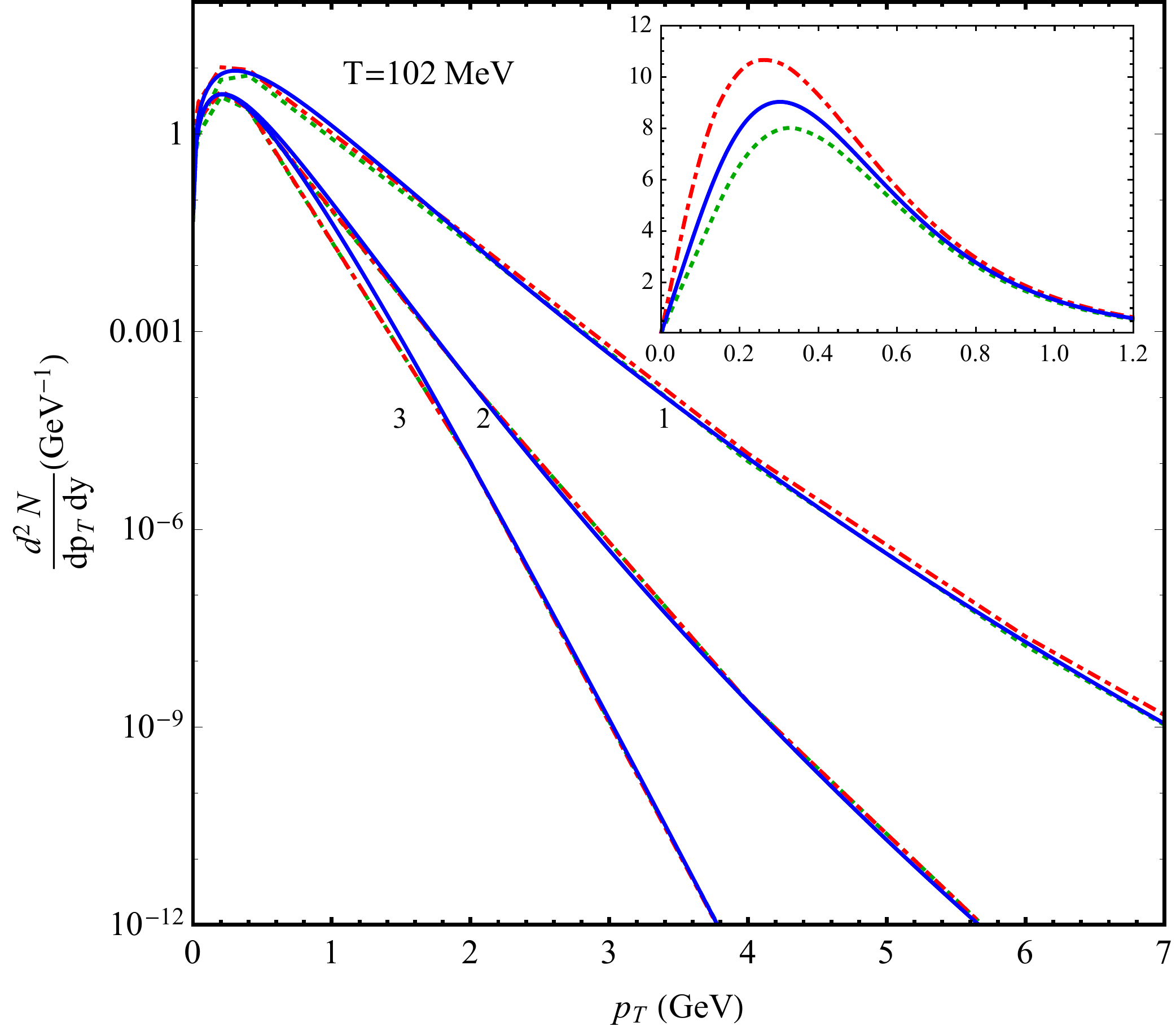}
\endminipage\hfill
\caption{(Color online) Comparison among the Maxwell-Boltzmann (blue solid line) , Fermi-Dirac (green dotted line) and Bose-Einstein (red dot-dashed line) spectra in the Tsallis-1 exact formalism at mid-rapidity $(y=0)$ for two different values of temperature. Temperature $T=82$ MeV (left panel) and $T=102$ MeV (right panel), chemical potential $\mu=0$, radius $R=4$ fm, the entropic parameter $q=0.98$ and mass $m=139.57$ MeV (pion mass). Curves $1,2$ and $3$ correspond to the exact Tsallis-1 statistics, zeroth-term approximation and the Boltzmann-Gibbs statistics $(q=1)$, respectively. Inset: the low transverse momentum variation of the spectra for the exact Tsallis-1 statistics.}
\label{fig6}
\end{figure*}

In Fig.~\ref{fig4}, we compare the mid-rapidity Maxwell - Boltzmann spectra in the Tsallis-1 and Tsallis-2 statistics for massive particles. We find that the spectra are different in the two approaches. This is starkly different from the massless case where the spectra calculated in the Tsallis-1 and Tsallis-2 overlap with each other~\cite{ParvanBaldin} under the multiplicative transformation $q\to 1/q$. The equivalence of the nonextensive formalisms under this replacement reflects the $q$-duality~\cite{Beck18}. Thus, we have found that for the Maxwell-Boltzmann statistics of particles the multiplicative $q$-duality of the Tsallis-1 and Tsallis-2 formalisms is preserved for massless particles and it is violated for massive ones. However, it is easy to verify that the multiplicative $q$-duality of the Tsallis-1 and Tsallis-2 formalisms in the zeroth term approximation is preserved for the Fermi-Dirac, Bose-Einstein and Maxwell-Boltzmann statistics of both massive and massless particles. For example, Eq.~(\ref{116}) for the the Fermi-Dirac, Bose-Einstein and Maxwell - Boltzmann statistics of particles of the Tsallis-2 statistics can be exactly obtained from Eq.~(\ref{65}) of the Tsallis-1 statistics by the replacement $q\to 1/q$.

\begin{figure}[!htb]
\begin{center}
\includegraphics[width=0.48\textwidth]{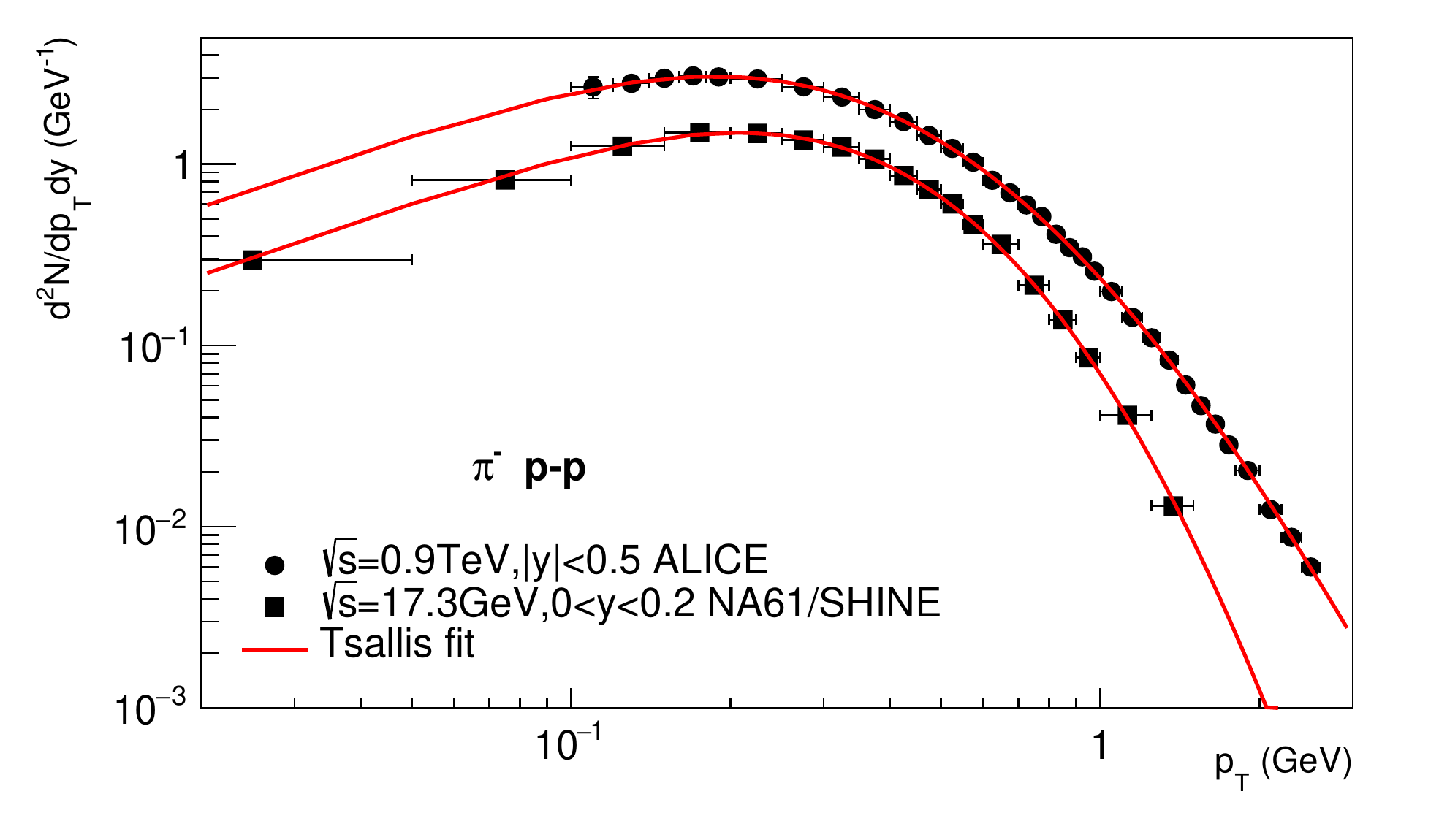}
\caption{(Color online) Transverse momentum distributions of $\pi^{-}$ produced in $pp$ collisions as obtained by the ALICE Collaboration at $\sqrt{s}=0.9$ TeV~\cite{ALICE_piplus} and the NA61/SHINE Collaboration at $\sqrt{s}=17.3$ GeV~\cite{NA61}. The solid curves are the fits of the data to the exact distribution function (\ref{50}) of the Tsallis-1 statistics for different values of the cut-off parameter $n_{0}$.    }
\label{fig7}
\end{center}
\end{figure}

\begin{table*}
\begin{tabular}{rrcccc}
 \hline
 \hline
 $\qquad$ $n_0$  & $\qquad$ $T$(MeV)$ \qquad$ & $\qquad$ $\qquad$ $R$(fm)$\qquad$ & $\qquad$ $\qquad$ $q$ $\qquad$ &$\qquad$$\chi^{2}/ndf$ $\qquad$ \\
 \hline
\hline
         0 & 71.837$\pm$2.423 & 4.743$\pm$0.134 & 0.873$\pm$0.004 & 2.214/30\\
         1 & 50.526$\pm$1.439 & 5.029$\pm$0.112 & 0.911$\pm$0.001 & 1.363/30\\
         2 & 42.245$\pm$0.233 & 5.553$\pm$0.032 & 0.9321$\pm$0.0003 & 1.827/30\\
         3 & 33.750$\pm$0.035 & 6.859$\pm$0.015 & 0.9442$\pm$0.0001 & 5.553/30\\
\hline
\hline
\end{tabular}
\caption{Parameters of the Tsallis-1 statistics fit with different cut-off $n_0$ for the $\pi^-$ particles produced in the $p$-$p$ collisions as obtained by the ALICE Collaboration at $\sqrt{s}$=0.9 TeV~\cite{ALICE_piplus}.}
\label{tabALICE}
\end{table*}

\begin{table*}
\begin{tabular}{rrcccc}
 \hline
 \hline
 $\qquad$ $n_0$  & $\qquad$ $T$(MeV)$ \qquad$ & $\qquad$ $\qquad$ $R$(fm)$\qquad$ & $\qquad$ $\qquad$ $q$ $\qquad$ &$\qquad$$\chi^{2}/ndf$ $\qquad$ \\
 \hline
\hline
    0 & 90.574$\pm$5.092 & 3.140$\pm$0.148 & 0.931$\pm$0.013 & 0.459/15\\
    1 & 79.224$\pm$8.610 & 3.145$\pm$0.225 & 0.946$\pm$0.011 & 0.453/15\\
    2 & 76.835$\pm$8.048 & 3.138$\pm$0.217 & 0.957$\pm$0.007 & 0.4503/15\\
    3 & 77.527$\pm$2.293 & 3.116$\pm$0.063 & 0.964$\pm$0.002 & 0.4575/15\\
\hline
\hline
\end{tabular}
\caption{Parameters of the Tsallis-1 statistics fit with different cut-off $n_0$ for the $\pi^-$ particles produced in the $p$-$p$ collisions as obtained by the NA61/SHINE Collaboration at $\sqrt{s}$=17.3 GeV\cite{NA61}.}
\label{tabNA61}
\end{table*}

Figures~\ref{fig5} and \ref{fig6} present the quantum statistics of particles. In Fig.~\ref{fig5}, a comparison of the mid-rapidity classical and quantum spectra of the Tsallis-1 (left panels) and the Tsallis-2 (right panels) statistics is displayed for different $q$ values. In this figure, the upper left and the lower left panels have $q$ values 0.88 and 0.98, respectively, whereas the upper right and the lower right panels have $q$ values 1.14 and 1.02 in that order. From the upper panels we observe that in the low momentum region the relative difference between the Fermi-Dirac (Bose-Einstein) and the Maxwell - Boltzmann spectra increases (decreases) and both of them become almost constant in the higher momentum region. However, in the spectra shown in the lower panels the difference between the Fermi-Dirac and the Maxwell-Boltzmann spectra grows more rapidly than that between the Bose-Einstein and the Maxwell-Boltzmann spectra. We plot the classical and quantum spectra in the Tsallis-1 statistics for two temperature values given by 82 MeV and 102 MeV in Fig.~\ref{fig6} and as expected, we see that at higher temperature the classical and quantum distributions tend to merge together. Hence, from the above discussion it is apparent that the quantum statistics becomes important at lower values of temperature and at higher $q$ values in the Tsallis-1 statistics, which correspond to lower $q$ values in the Tsallis-2 statistics. This region of low temperature
and low $q$ (in Tsallis-2 where $q>1$) values is probed by the experiments like NA61/SHINE and will be probed in the forthcoming accelerators like NICA~\cite{Parvan16,Parvan17a}. Hence, we infer that the description of the transverse momentum distribution of hadrons in terms of the classical Tsallis statistics will be insufficient because one needs to evoke the quantum statistics for the particles produced in this regime.

Figure~\ref{fig7} represents the comparison of the theoretical transverse momentum distribution of the Tsallis-1 statistics with the experimental data for the $\pi^-$ particles obtained by the ALICE and the NA61/SHINE collaborations in the proton-proton collisions at the center of mass energies 0.9 TeV and 17.3 GeV, respectively. We vary the cut-off parameter $n_0$ which defines the number of terms included in the series expansion that represents the hadron transverse momentum spectrum in Eq.~(\ref{50}). The fitting parameters $q$, $T$, and $R$ (where $R$ signifies the radius of a spherical volume $V$) with varying $n_0$ are tabulated in Tables \ref{tabALICE} and \ref{tabNA61}. For both energies we observe that an increase in the number of terms in the series results in an increase of the central value of the $q$ parameter. For ALICE data, the value of temperature decreases, whereas there is no such monotonicity for NA61/SHINE data. The fitted value of radius increases with $n_0$ for ALICE data, but it remains more or less fixed within the error bars for NA61/SHINE data. We also observe that the theoretical curves overlap with each other for different values of the cut-off parameter $n_{0}$. As it is seen from Fig.~\ref{fig7}, the exact distribution function (\ref{50}) for the massive particles of the Tsallis-1 statistics very well describes  the experimental data for the low $p_{T}$ momenta of hadrons measured by the NA61/SHINE Collaboration at $\sqrt{s}$=17.3 GeV. In contrast to this, the exact distribution function for massless particles of the Tsallis-1 statistics does not describe well the same experimental data in the region of low $p_{T}$ momenta (see Fig.1 in Ref.~\cite{Parvan16}). This comparison proves that the consideration of the mass of pions in the Tsallis formalism is important for the description of the transverse momentum spectra of low $p_{T}$ pions at low collision energies.

\section{Summary, conclusions and outlook}\label{sec5}
To summarize, we have obtained the exact analytical formulae for the transverse momentum distributions of the relativistic massive particles following the Bose-Einstein, Fermi-Dirac and Maxwell-Boltzmann statistics in the grand canonical ensemble in the framework of both the Tsallis normalized and the Tsallis unnormalized statistics. The exact analytical results were expressed in a general form as a series expansion in the integral representation. In particular, for the classical Maxwell-Boltzmann statistics of particles, the terms of the series expansion can be written explicitly in the form of integrals of the integer powers of the modified Bessel functions. We have also found the analytical formulae for the transverse momentum distributions in the zeroth term approximation for all three statistics of particles (Bose-Einstein, Fermi-Dirac and Maxwell-Boltzmann). We have revealed that the phenomenological Tsallis distribution for the Maxwell-Boltzmann statistics of particles proposed in Ref.~\cite{Cleymans2012} (which is largely used in high energy physics) exactly coincides with the transverse momentum distribution of the Tsallis unnormalized statistics in the zeroth term approximation. Moreover, we have also found that this phenomenological Tsallis distribution can be obtained from the zeroth term approximation of the Tsallis normalized statistics by the transformation of the entropic parameter $q\to 1/q$. However, we have shown that the quantum phenomenological Tsallis distributions proposed in Ref.~\cite{Cleymans2012} do not correspond to the exact or zeroth term approximate results of both the Tsallis normalized and Tsallis unnormalized statistics. It should be stressed that the same results were analytically obtained earlier in Ref.~\cite{Parvan17} by one of the authors of this paper for the Maxwell-Boltzmann statistics of massless particles and hence, the present work stands as a generalization of that to the quantum statistics of particles and nonzero particle mass. In the massless ($m\rightarrow 0$) limit the calculations of the present paper for the classical statistics of particles resemble the results of the previous work.

We have calculated numerically the transverse momentum distributions of massive hadrons. We have revealed that in the ranges of the entropic parameter relevant to the processes of high energy physics ($q<1$ for Tsallis-1 and $q>1$ for Tsallis-2) the exact Tsallis statistics is divergent. The Tsallis statistics converges for $q=1$ when it recovers the Boltzmann-Gibbs statistics. Nevertheless, we have found that there is a few terms in the series expansions in the vicinity of the zeroth order term which are finite and have physical meaning in the processes of high energy physics. The number of these physical terms depends on the value of $q$ and increases infinitely with $q\to 1$. Therefore, to obtain physical results we have regularized the Tsallis statistics by introducing the upper cut-off in the series expansions.

We have compared the numerical results of the exact transverse momentum distribution of the Tsallis statistics with the phenomenological Tsallis distribution. We have revealed that the exact results of the Tsallis statistics (Tsallis-1 and Tsallis-2) strongly enhance the production of the high-$p_{T}$ hadrons in the transverse momentum distributions in comparison with the phenomenological Tsallis distribution function at the same values of $q$. We have also found that in contrast to the case of massless particles, the $q$-duality of the Tsallis normalized and Tsallis unnormalized statistics in the case of massive particles is not followed. However, the $q$-duality for both these two nonextensive formalisms in the zeroth term approximation is preserved for both massive and massless particles and for all three particle statistics.

The exact numerical results for the quantum statistics of particles (Bose-Einstein and Fermi-Dirac) in the Tsallis normalized and Tsallis unnormalized statistics of microstates have been obtained. We have shown that the quantum statistics of particles in the Tsallis distributions is significant at small values of the transverse momentum of hadrons up to $1$ GeV and at small energies of heavy-ion or $pp$ collisions.

We have compared the exact transverse momentum distribution of the Tsallis normalized statistics for the Maxwell-Boltzmann statistics of massive particles with the experimental data for the negatively charged pions obtained by the ALICE and the NA61/SHINE collaborations in the proton-proton collisions. We have shown that in comparison with the massless Tsallis distribution the consideration of the mass of pions in the Tsallis formalism sufficiently improves the description of the transverse momentum spectra of low $p_{T}$ pions at low collision energies.

For completeness of the present research we have derived in a general form the equilibrium probability distribution functions of microstates of the system for both the Tsallis normalized and unnormalized statistics in the grand canonical ensemble from the second law of thermodynamics. The probabilities of microstates and thermodynamic quantities for both nonextensive statistics formalisms were expressed analytically in terms of the corresponding Boltzmann-Gibbs quantities using the integral representation. The exact results for the relativistic ideal gas of massive particles following the Bose-Einstein, Fermi-Dirac and Maxwell-Boltzmann statistics in the grand canonical ensemble in the framework of both the Tsallis normalized and the Tsallis unnormalized statistics were also provided.

The results of the present paper allow to reanalyze the experimental data on the transverse momentum spectra of particles produced in proton-proton and heavy-ion collisions measured by such collaborations as NA61/SHINE, PHENIX, ALICE to study the effect of the exact Tsallis statistics. These results can also be compared with the earlier works which have utilized the phenomenological Tsallis distributions. In addition to that, the quantum spectra calculated in this paper can also be used for the analysis of the experimental data. And finally, the same calculations can be performed for other formalisms of the Tsallis statistics (e.g. the Tsallis-3 formalism) and can be compared/contrasted with the present work and the results of the experiments.

The importance of our work is that it orders the multitude of phenomenological $q$-functions and shows that there is only one unique transverse momentum distribution function belonging to the Tsallis statistics (in the form of the series expansion or in the form of integral). This means that all other phenomenological Tsallis distributions (largely used in high-energy physics) either do not belong to the Tsallis statistics at all or have an indirect relation in the form of the approximation to an incorrect variant of the Tsallis statistics, namely, to the Tsallis-2 statistics. The present work shows exactly which transverse momentum distribution function belongs to the Tsallis statistics and which function should be used to analyze the experimental data if we deal with the Tsallis statistics. Thus, the results of the present paper prove that the phenomenological Tsallis distributions provide the incorrect results for the thermodynamic quantities extracted from the experimental data as these distributions do not recover the unique form of the single-particle distribution function of the Tsallis statistics and thus lack the correct link with the second law of thermodynamics. This is the sense of our work. The results of our paper, as we know, have not been reported by any other authors anywhere else so far.


\begin{acknowledgement}
This work was supported in part by the joint research project of JINR and IFIN-HH. We are indebted to D.-V.~Anghel, J.~Cleymans, G.I.~Lykasov, A.S.~Sorin and O.V.~Teryaev for stimulating discussions. We also grateful to S.~Grigoryan for valuable remarks.
\end{acknowledgement}

\appendix

\section{Ideal gas in the Tsallis-1 statistics}\label{ap1}
Let us consider the ideal gas for the Tsallis normalized statistics in the grand canonical ensemble for both quantum and classical statistics of particles and find the transverse momentum distribution of hadrons.

\subsection{Exact results}
\subsubsection{General case}
The norm equation (\ref{8}) for the ideal gas in the occupation number representation can be written as~\cite{Parvan17}
\begin{equation}\label{20}
  \sum\limits_{\{n_{\mathbf{p}\sigma}\}}  G\{n_{\mathbf{p}\sigma}\}   \left[1+\frac{q-1}{q}\frac{\Lambda- \sum\limits_{\mathbf{p},\sigma} n_{\mathbf{p}\sigma} (\varepsilon_{\mathbf{p}}-\mu)}{T}\right]^{\frac{1}{q-1}} = 1,
\end{equation}
where $n_{\mathbf{p}\sigma}=0,1,\ldots,K$ are the occupation numbers, \\ $G\{n_{\mathbf{p}\sigma}\} = 1$ for the Fermi-Dirac $(K=1)$ and Boze-Einstein $(K=\infty)$ statistics of particles, and $G\{n_{\mathbf{p}\sigma}\}=1/(\prod_{\mathbf{p}\sigma}n_{\mathbf{p}\sigma}!)$ for the Maxwell-Boltzmann $(K=\infty)$ statistics of particles. Using Eqs.~(\ref{20}), (\ref{14}) and (\ref{15}), we obtain Eqs.~(\ref{21}) and (\ref{22}) with $\Omega_{G}$ calculated by Eq.~(\ref{23}).

The mean occupation numbers for the ideal gas of the Tsallis normalized statistics in the grand canonical ensemble can be defined as~\cite{Parvan17,ParvanBaldin}
\begin{eqnarray}\label{24}
  \langle n_{\mathbf{p}\sigma} \rangle &=& \sum\limits_{\{n_{\mathbf{p}\sigma}\}} n_{\mathbf{p}\sigma} G\{n_{\mathbf{p}\sigma}\} \nonumber \\  &\times&  \left[1+\frac{q-1}{q}\frac{\Lambda- \sum\limits_{\mathbf{p},\sigma} n_{\mathbf{p}\sigma} (\varepsilon_{\mathbf{p}}-\mu)}{T}\right]^{\frac{1}{q-1}}.
\end{eqnarray}
The mean occupation numbers (\ref{24}) for the Tsallis normalized statistics in the integral representation (\ref{17})--(\ref{19}) take the forms
\begin{eqnarray}\label{25}
 \langle n_{\mathbf{p}\sigma} \rangle &=& \sum\limits_{n=0}^{\infty}  \frac{1}{n!\Gamma\left(\frac{1}{1-q}\right)} \int\limits_{0}^{\infty} t^{\frac{q}{1-q}} e^{-t\left[1+\frac{q-1}{q}\frac{\Lambda}{T}\right]} \nonumber \\ &\times& \left(-\beta'\Omega_{G}\left(\beta'\right)\right)^{n}  \langle n_{\mathbf{p}\sigma} \rangle_{G}\left(\beta'\right) dt \quad \mathrm{for}
 \quad q<1 \;\;\;\;\;\;\;
\end{eqnarray}
and
\begin{eqnarray}\label{26}
 \langle n_{\mathbf{p}\sigma} \rangle &=& \sum\limits_{n=0}^{\infty} \frac{\Gamma\left(\frac{q}{q-1}\right)}{n!}    \frac{i}{2\pi} \oint\limits_{C} (-t)^{\frac{q}{1-q}} e^{-t\left[1+\frac{q-1}{q}\frac{\Lambda}{T}\right]} \nonumber \\ &\times& \left(-\beta'\Omega_{G}\left(\beta'\right)\right)^{n}  \langle n_{\mathbf{p}\sigma} \rangle_{G}\left(\beta'\right) dt \quad \mathrm{for} \quad q>1, \;\;\;\;\;\;\;
\end{eqnarray}
where
\begin{equation}\label{27}
  \langle n_{\mathbf{p}\sigma} \rangle_{G}\left(\beta'\right) = \frac{1}{e^{\beta' (\varepsilon_{\mathbf{p}}-\mu)}+\eta}
\end{equation}
are the mean occupation numbers of the ideal gas of the Boltzmann-Gibbs statistics.

The mean energy and the mean number of particles of the system for the ideal gas of the Tsallis normalized statistics in the grand canonical ensemble can be written as~\cite{Parvan17}
\begin{eqnarray}\label{28}
  \langle H \rangle &=& \sum\limits_{\{n_{\mathbf{p}\sigma}\}}   G\{n_{\mathbf{p}\sigma}\}  \left(\sum\limits_{\mathbf{p},\sigma} n_{\mathbf{p}\sigma}\varepsilon_{\mathbf{p}}\right) \nonumber \\ &\times& \left[1+\frac{q-1}{q}\frac{\Lambda- \sum\limits_{\mathbf{p},\sigma} n_{\mathbf{p}\sigma} (\varepsilon_{\mathbf{p}}-\mu)}{T}\right]^{\frac{1}{q-1}} \nonumber \\
  &=& \sum\limits_{\mathbf{p},\sigma} \langle n_{\mathbf{p}\sigma} \rangle \varepsilon_{\mathbf{p}}
\end{eqnarray}
and
\begin{eqnarray}\label{29}
  \langle N \rangle &=& \sum\limits_{\{n_{\mathbf{p}\sigma}\}}   G\{n_{\mathbf{p}\sigma}\}  \left(\sum\limits_{\mathbf{p},\sigma} n_{\mathbf{p}\sigma}\right) \nonumber \\ &\times&  \left[1+\frac{q-1}{q}\frac{\Lambda- \sum\limits_{\mathbf{p},\sigma} n_{\mathbf{p}\sigma} (\varepsilon_{\mathbf{p}}-\mu)}{T}\right]^{\frac{1}{q-1}} \nonumber \\ &=& \sum\limits_{\mathbf{p},\sigma} \langle n_{\mathbf{p}\sigma} \rangle,
\end{eqnarray}
where the mean occupation numbers $\langle n_{\mathbf{p}\sigma} \rangle$ are calculated by Eqs.~(\ref{25})--(\ref{27}). Then, the mean energy (\ref{28}) and the mean number of particles (\ref{29}) of the system for the Tsallis normalized statistics in the integral representation (\ref{17})--(\ref{19}) are
\begin{eqnarray}\label{30}
 \langle H \rangle &=& \sum\limits_{n=0}^{\infty}  \frac{1}{n!\Gamma\left(\frac{1}{1-q}\right)} \int\limits_{0}^{\infty} t^{\frac{q}{1-q}} e^{-t\left[1+\frac{q-1}{q}\frac{\Lambda}{T}\right]} \nonumber \\ &\times&   \left(-\beta'\Omega_{G}\left(\beta'\right)\right)^{n}  \langle H \rangle_{G}\left(\beta'\right) dt \quad \mathrm{for} \quad q<1  \\ \label{31}
 \langle H \rangle &=& \sum\limits_{n=0}^{\infty} \frac{\Gamma\left(\frac{q}{q-1}\right)}{n!}    \frac{i}{2\pi} \oint\limits_{C} (-t)^{\frac{q}{1-q}} e^{-t\left[1+\frac{q-1}{q}\frac{\Lambda}{T}\right]} \nonumber \\ &\times&  \left(-\beta'\Omega_{G}\left(\beta'\right)\right)^{n}  \langle H \rangle_{G}\left(\beta'\right) dt \quad \mathrm{for} \quad q>1 \\ \label{32}
\langle N \rangle &=& \sum\limits_{n=0}^{\infty}  \frac{1}{n!\Gamma\left(\frac{1}{1-q}\right)} \int\limits_{0}^{\infty} t^{\frac{q}{1-q}} e^{-t\left[1+\frac{q-1}{q}\frac{\Lambda}{T}\right]} \nonumber \\ &\times&  \left(-\beta'\Omega_{G}\left(\beta'\right)\right)^{n}  \langle N \rangle_{G}\left(\beta'\right) dt \quad \mathrm{for} \quad q<1 \\ \label{33}
 \langle N \rangle &=& \sum\limits_{n=0}^{\infty} \frac{\Gamma\left(\frac{q}{q-1}\right)}{n!}    \frac{i}{2\pi} \oint\limits_{C} (-t)^{\frac{q}{1-q}} e^{-t\left[1+\frac{q-1}{q}\frac{\Lambda}{T}\right]} \nonumber \\ &\times&  \left(-\beta'\Omega_{G}\left(\beta'\right)\right)^{n}  \langle N \rangle_{G}\left(\beta'\right) dt \quad \mathrm{for} \quad q>1, \;\;\;\;\;
\end{eqnarray}
where
\begin{eqnarray}\label{34a}
  \langle H \rangle_{G}\left(\beta'\right) &=& \sum\limits_{\mathbf{p},\sigma} \frac{\varepsilon_{\mathbf{p}}}{e^{\beta' (\varepsilon_{\mathbf{p}}-\mu)}+\eta} \\ \label{34b}
  \langle N \rangle_{G}\left(\beta'\right) &=& \sum\limits_{\mathbf{p},\sigma} \frac{1}{e^{\beta' (\varepsilon_{\mathbf{p}}-\mu)}+\eta}
\end{eqnarray}
are the mean energy and the mean number of particles, respectively, for the ideal gas of the Boltzmann-Gibbs statistics.

The transverse momentum distribution of particles is related to the mean occupation numbers as
\begin{equation}\label{35}
  \frac{d^{2}N}{dp_{T}dy} = \frac{V}{(2\pi)^{3}} \int\limits_{0}^{2\pi} d\varphi p_{T} \varepsilon_{\mathbf{p}} \ \sum\limits_{\sigma} \langle n_{\mathbf{p}\sigma}\rangle,
\end{equation}
where $\varepsilon_{\mathbf{p}}=m_{T} \cosh y$. Using Eqs.~(\ref{25}) and (\ref{26}), we obtain Eqs.~(\ref{36}) and (\ref{37}).

\subsubsection{Maxwell-Boltzmann statistics of particles}
Let us derive in more detail the formulas for the Maxwell-Boltzmann statistics of particles. Substituting Eqs.~(\ref{38}) and (\ref{27}) for $\eta=0$ into Eqs.~(\ref{25}) and (\ref{26}), we find the mean occupation numbers for the Maxwell - Boltzmann  statistics of particles as
\begin{eqnarray}\label{42}
 \langle n_{\mathbf{p}\sigma} \rangle &=& \sum\limits_{n=0}^{\infty} \frac{\omega^{n}}{n!} \frac{1}{\Gamma\left(\frac{1}{1-q}\right)} \int\limits_{0}^{\infty} t^{\frac{q}{1-q}-n} \nonumber \\ &\times&
 e^{-t\left[1+\frac{q-1}{q}\frac{\Lambda-\varepsilon_{\mathbf{p}}+\mu(n+1)}{T}\right]} \nonumber \\ &\times&
 \left(K_{2}\left(\frac{t(1-q)m}{qT} \right)\right)^{n} dt \quad \mathrm{for} \quad q<1
\end{eqnarray}
and
\begin{eqnarray}\label{43}
 \langle n_{\mathbf{p}\sigma} \rangle &=& \sum\limits_{n=0}^{\infty} \frac{(-\omega)^{n}}{n!}  \Gamma\left(\frac{q}{q-1}\right)  \frac{i}{2\pi} \oint\limits_{C} (-t)^{\frac{q}{1-q}-n} \nonumber \\ &\times& e^{-t\left[1+\frac{q-1}{q}\frac{\Lambda-\varepsilon_{\mathbf{p}}+\mu(n+1)}{T}\right]} \nonumber \\ &\times&  \left(K_{2}\left(\frac{t(1-q)m}{qT} \right)\right)^{n} dt \quad \mathrm{for} \quad q>1.
\end{eqnarray}
In the ultrarelativistic limit $(m=0)$, Eqs.~(\ref{42}) and (\ref{43}) recover Eq.~(25)of Ref.~\cite{Parvan17}.

The mean number of particles of the Boltzmann-Gibbs statistics (\ref{34b}) in the case of the Maxwell-Boltzmann statistics of particles can be written as
\begin{equation}\label{44}
  \langle N \rangle_{G}\left(\beta'\right)=  \frac{gV}{2\pi^{2}} \frac{m^{2}}{\beta'} e^{\beta'\mu} K_{2}\left(\beta'm\right).
\end{equation}
Substituting Eqs.~(\ref{38}) and (\ref{44}) into Eqs.~(\ref{32}) and (\ref{33}), we obtain the mean number of particles for the Maxwell-Boltzmann statistics as
\begin{eqnarray}\label{45}
 \langle N \rangle &=& \sum\limits_{n=0}^{\infty} \frac{\omega^{n+1}}{n!} \frac{1}{\Gamma\left(\frac{1}{1-q}\right)} \int\limits_{0}^{\infty} t^{\frac{q}{1-q}-n-1} \nonumber \\ &\times&
 e^{-t\left[1+\frac{q-1}{q}\frac{\Lambda+\mu(n+1)}{T}\right]} \nonumber \\ &\times& \left(K_{2}\left(\frac{t(1-q)m}{qT} \right)\right)^{n+1} dt \quad \mathrm{for} \quad q<1
\end{eqnarray}
and
\begin{eqnarray}\label{46}
 \langle N \rangle &=& \sum\limits_{n=0}^{\infty} \frac{(-\omega)^{n+1}}{n!}  \Gamma\left(\frac{q}{q-1}\right)  \frac{i}{2\pi} \oint\limits_{C} (-t)^{\frac{q}{1-q}-n-1} \nonumber \\ &\times& e^{-t\left[1+\frac{q-1}{q}\frac{\Lambda+\mu(n+1)}{T}\right]} \nonumber \\ &\times& \left(K_{2}\left(\frac{t(1-q)m}{qT} \right)\right)^{n+1} dt \quad \mathrm{for} \quad q>1.
\end{eqnarray}
In the ultrarelativistic limit $(m=0)$, Eqs.~(\ref{45}) and (\ref{46}) recover Eq.~(27) of Ref.~\cite{Parvan17}.

The mean energy of the Boltzmann-Gibbs statistics (\ref{34a}) in the case of the Maxwell-Boltzmann statistics of particles can be written as
\begin{equation}\label{47}
  \langle H \rangle_{G}\left(\beta'\right)=  \frac{gV}{2\pi^{2}} \frac{m^{3}}{\beta'} e^{\beta'\mu} \left[K_{3}\left(\beta'm\right) - \frac{K_{2}\left(\beta'm\right)}{\beta'm} \right].
\end{equation}
Substituting Eqs.~(\ref{38}) and (\ref{47}) into Eqs.~(\ref{30}) and (\ref{31}), we obtain the mean energy of the system for the Maxwell-Boltzmann statistics of particles as
\begin{eqnarray}\label{48}
 \langle H \rangle &=& \sum\limits_{n=0}^{\infty} \frac{\omega^{n+1}}{n!} \frac{1}{\Gamma\left(\frac{1}{1-q}\right)} \int\limits_{0}^{\infty} t^{\frac{q}{1-q}-n-1} \nonumber \\ &\times&
 e^{-t\left[1+\frac{q-1}{q}\frac{\Lambda+\mu(n+1)}{T}\right]} \left(K_{2}\left(\frac{t(1-q)m}{qT} \right)\right)^{n+1} \nonumber \\
 &\times& \left[m\frac{K_{3}\left(\frac{t(1-q)m}{qT} \right)}{K_{2}\left(\frac{t(1-q)m}{qT} \right)}- t^{-1}\frac{qT}{1-q}  \right] dt \nonumber \\ && \quad \mathrm{for} \quad q<1
\end{eqnarray}
and
\begin{eqnarray}\label{49}
 \langle H \rangle &=& \sum\limits_{n=0}^{\infty} \frac{(-\omega)^{n+1}}{n!}  \Gamma\left(\frac{q}{q-1}\right)  \frac{i}{2\pi} \oint\limits_{C} (-t)^{\frac{q}{1-q}-n-1} \nonumber \\ &\times& e^{-t\left[1+\frac{q-1}{q}\frac{\Lambda+\mu(n+1)}{T}\right]} \left(K_{2}\left(\frac{t(1-q)m}{qT} \right)\right)^{n+1} \nonumber \\
 &\times& \left[m\frac{K_{3}\left(\frac{t(1-q)m}{qT} \right)}{K_{2}\left(\frac{t(1-q)m}{qT} \right)}- t^{-1}\frac{qT}{1-q}  \right] dt \nonumber \\ && \quad \mathrm{for} \quad q>1.
\end{eqnarray}
In the ultrarelativistic limit $(m=0)$, Eqs.~(\ref{48}) and (\ref{49}) recover Eq.~(29) of Ref.~\cite{Parvan17}.

Substituting Eqs.~(\ref{42}) and (\ref{43}) into Eq.~(\ref{35}), we obtain Eqs.~(\ref{50}) and (\ref{51}) for the transverse momentum distribution of the Maxwell-Boltzmann statistics of particles in the Tsallis normalized statistics.

\subsection{Zeroth term approximation}
Substituting $\Lambda=0$ into Eqs.~(\ref{25}) and (\ref{26}) and considering only the zeroth term, we obtain
\begin{equation}\label{52}
 \langle n_{\mathbf{p}\sigma} \rangle =  \frac{1}{\Gamma\left(\frac{1}{1-q}\right)} \int\limits_{0}^{\infty} t^{\frac{q}{1-q}} e^{-t} \frac{1}{e^{\beta' (\varepsilon_{\mathbf{p}}-\mu)}+\eta}  dt \quad \mathrm{for} \quad q<1
\end{equation}
and
\begin{eqnarray}\label{53}
 \langle n_{\mathbf{p}\sigma} \rangle &=& \Gamma\left(\frac{q}{q-1}\right)  \frac{i}{2\pi} \oint\limits_{C} (-t)^{\frac{q}{1-q}} e^{-t}  \frac{1}{e^{\beta' (\varepsilon_{\mathbf{p}}-\mu)}+\eta} dt \nonumber \\ && \mathrm{for} \quad q>1.
\end{eqnarray}
Using Eqs.~(\ref{10}), (\ref{11}) and (\ref{54}), we obtain the mean occupation numbers in the zeroth term approximation as
\begin{eqnarray}\label{56}
\langle n_{\mathbf{p}\sigma} \rangle &=& \sum\limits_{k=0}^{\infty} (-\eta)^{k} \left[1+(k+1) \frac{1-q}{q} \frac{\varepsilon_{\mathbf{p}}-\mu}{T} \right]^{\frac{1}{q-1}} \nonumber \\ && \mathrm{for} \quad \eta=-1,0,1.
\end{eqnarray}
In the ultrarelativistic limit $(m=0)$, Eq.~(\ref{56}) recovers Eq.~(37) of Ref.~\cite{Parvan17}. Substituting Eq.~(\ref{56}) into Eq.~(\ref{28}), we obtain the mean energy of the system in the zeroth term approximation as
\begin{eqnarray}\label{59}
\langle H \rangle &=& \sum\limits_{k=0}^{\infty} (-\eta)^{k} \sum\limits_{\mathbf{p},\sigma} \varepsilon_{\mathbf{p}} \left[1+(k+1) \frac{1-q}{q} \frac{\varepsilon_{\mathbf{p}}-\mu}{T} \right]^{\frac{1}{q-1}} \nonumber \\ && \mathrm{for} \quad \eta=-1,0,1. \;\;\;\;
\end{eqnarray}
Now substituting Eq.~(\ref{56}) into Eq.~(\ref{29}), we obtain the mean number of particles of the system in the zeroth term approximation as
\begin{eqnarray}\label{62}
\langle N \rangle &=& \sum\limits_{k=0}^{\infty} (-\eta)^{k} \sum\limits_{\mathbf{p},\sigma} \left[1+(k+1) \frac{1-q}{q} \frac{\varepsilon_{\mathbf{p}}-\mu}{T} \right]^{\frac{1}{q-1}} \nonumber \\ && \mathrm{for} \quad \eta=-1,0,1.
\end{eqnarray}
Substituting Eq.~(\ref{56}) into Eq.~(\ref{35}), we obtain Eqs.~(\ref{65}) and (\ref{66}) for the transverse momentum distribution.

\section{Ideal gas in the Tsallis-2 statistics}\label{ap2}
Let us consider the ideal gas of hadrons for the Tsallis unnormalized statistics in the grand canonical ensemble for both quantum and classical statistics of particles and calculate the transverse momentum distribution of hadrons.

\subsection{Exact results}
\subsubsection{General case}
The partition function (\ref{71}) for the ideal gas in the occupation number representation is~\cite{Parvan17,ParvanBaldin}
\begin{equation}\label{79}
  Z =\sum\limits_{\{n_{\mathbf{p}\sigma}\}}  G\{n_{\mathbf{p}\sigma}\}   \left[1-(1-q)\frac{\sum\limits_{\mathbf{p},\sigma} n_{\mathbf{p}\sigma} (\varepsilon_{\mathbf{p}}-\mu)}{T}\right]^{\frac{1}{1-q}}.
\end{equation}
Then, the partition function (\ref{79}) in the integral representation (\ref{75}) and (\ref{76}) is represented by Eqs.~(\ref{80}) and (\ref{81}).

The mean occupation numbers for the ideal gas of the Tsallis unnormalized statistics in the grand canonical ensemble can be defined as~\cite{Parvan17,ParvanBaldin}
\begin{eqnarray}\label{82}
  \langle n_{\mathbf{p}\sigma} \rangle &=& \frac{1}{Z^{q}} \sum\limits_{\{n_{\mathbf{p}\sigma}\}} n_{\mathbf{p}\sigma} G\{n_{\mathbf{p}\sigma}\} \nonumber \\ &\times&  \left[1-(1-q)\frac{\sum\limits_{\mathbf{p},\sigma} n_{\mathbf{p}\sigma} (\varepsilon_{\mathbf{p}}-\mu)}{T}\right]^{\frac{q}{1-q}}.
\end{eqnarray}
The mean occupation numbers (\ref{82}) for the Tsallis unnormalized statistics in the integral representation (\ref{77}), (\ref{78}) take the form
\begin{eqnarray}\label{83}
 \langle n_{\mathbf{p}\sigma} \rangle &=& \sum\limits_{n=0}^{\infty} \frac{1}{n!Z^{q}\Gamma\left(\frac{q}{q-1}\right)} \int\limits_{0}^{\infty} t^{\frac{1}{q-1}} e^{-t} \nonumber \\ &\times& \left(-\beta'\Omega_{G}\left(\beta'\right)\right)^{n}  \langle n_{\mathbf{p}\sigma} \rangle_{G}\left(\beta'\right) dt \quad \mathrm{for} \quad q>1 \;\;\;\;\;\;
\end{eqnarray}
and
\begin{eqnarray}\label{84}
 \langle n_{\mathbf{p}\sigma} \rangle &=& \sum\limits_{n=0}^{\infty}  \frac{\Gamma\left(\frac{1}{1-q}\right)}{n!Z^{q}}    \frac{i}{2\pi} \oint\limits_{C} (-t)^{\frac{1}{q-1}} e^{-t} \nonumber \\ &\times& \left(-\beta'\Omega_{G}\left(\beta'\right)\right)^{n}  \langle n_{\mathbf{p}\sigma} \rangle_{G}\left(\beta'\right) dt \quad \mathrm{for} \quad q<1, \;\;\;\;\;\;
\end{eqnarray}
where $\langle n_{\mathbf{p}\sigma} \rangle_{G}\left(\beta'\right)$ is defined in Eq.~(\ref{27}).

The mean energy and the mean number of particles of the system for the ideal gas of the Tsallis unnormalized statistics in the grand canonical ensemble can be written as
\begin{eqnarray}\label{85}
  \langle H \rangle &=& \frac{1}{Z^{q}}\sum\limits_{\{n_{\mathbf{p}\sigma}\}}   G\{n_{\mathbf{p}\sigma}\}  \left(\sum\limits_{\mathbf{p},\sigma} n_{\mathbf{p}\sigma}\varepsilon_{\mathbf{p}}\right) \nonumber \\ &\times& \left[1-(1-q)\frac{\sum\limits_{\mathbf{p},\sigma} n_{\mathbf{p}\sigma} (\varepsilon_{\mathbf{p}}-\mu)}{T}\right]^{\frac{q}{1-q}} \nonumber \\ &=& \sum\limits_{\mathbf{p},\sigma} \langle n_{\mathbf{p}\sigma} \rangle \varepsilon_{\mathbf{p}}
\end{eqnarray}
and
\begin{eqnarray}\label{86}
  \langle N \rangle &=&  \frac{1}{Z^{q}} \sum\limits_{\{n_{\mathbf{p}\sigma}\}}   G\{n_{\mathbf{p}\sigma}\}  \left(\sum\limits_{\mathbf{p},\sigma} n_{\mathbf{p}\sigma}\right) \nonumber \\ &\times& \left[1-(1-q)\frac{\sum\limits_{\mathbf{p},\sigma} n_{\mathbf{p}\sigma} (\varepsilon_{\mathbf{p}}-\mu)}{T}\right]^{\frac{q}{1-q}} \nonumber \\ &=& \sum\limits_{\mathbf{p},\sigma} \langle n_{\mathbf{p}\sigma} \rangle,
\end{eqnarray}
where the mean occupation numbers $\langle n_{\mathbf{p}\sigma} \rangle$ can be calculated by Eqs.~(\ref{83})--(\ref{84}). The mean energy (\ref{85}) and the mean number of particles (\ref{86}) of the system for the Tsallis unnormalized statistics in the integral representation (\ref{77}), (\ref{78}) are
\begin{eqnarray}\label{87}
 \langle H \rangle &=& \sum\limits_{n=0}^{\infty} \frac{1}{n!Z^{q}\Gamma\left(\frac{q}{q-1}\right)} \int\limits_{0}^{\infty} t^{\frac{1}{q-1}} e^{-t} \nonumber \\ &\times& \left(-\beta'\Omega_{G}\left(\beta'\right)\right)^{n}  \langle H \rangle_{G}\left(\beta'\right) dt \quad \mathrm{for} \quad q>1  \\ \label{88}
 \langle H \rangle &=& \sum\limits_{n=0}^{\infty}  \frac{\Gamma\left(\frac{1}{1-q}\right)}{n!Z^{q}}   \frac{i}{2\pi} \oint\limits_{C} (-t)^{\frac{1}{q-1}} e^{-t} \nonumber \\ &\times& \left(-\beta'\Omega_{G}\left(\beta'\right)\right)^{n}  \langle H \rangle_{G}\left(\beta'\right) dt \quad \mathrm{for} \quad q<1 \\ \label{89}
\langle N \rangle &=& \sum\limits_{n=0}^{\infty}  \frac{1}{n!Z^{q}\Gamma\left(\frac{q}{q-1}\right)} \int\limits_{0}^{\infty} t^{\frac{1}{q-1}} e^{-t} \nonumber \\ &\times& \left(-\beta'\Omega_{G}\left(\beta'\right)\right)^{n}  \langle N \rangle_{G}\left(\beta'\right) dt \quad \mathrm{for} \quad q>1 \\ \label{90}
 \langle N \rangle &=& \sum\limits_{n=0}^{\infty} \frac{\Gamma\left(\frac{1}{1-q}\right)}{n!Z^{q}}    \frac{i}{2\pi} \oint\limits_{C} (-t)^{\frac{1}{q-1}} e^{-t} \nonumber \\ &\times& \left(-\beta'\Omega_{G}\left(\beta'\right)\right)^{n}  \langle N \rangle_{G}\left(\beta'\right) dt \quad \mathrm{for} \quad q<1,
\end{eqnarray}
where $\langle H \rangle_{G}\left(\beta'\right)$ and $\langle N \rangle_{G}\left(\beta'\right)$ are defined in Eqs.~(\ref{34a}) and (\ref{34b}), respectively.

Substituting Eqs.~(\ref{83}) and (\ref{84}) into Eq.~(\ref{35}), we obtain Eqs.~(\ref{91}) and (\ref{92}).

\subsubsection{Maxwell-Boltzmann statistics of particles}
Substituting Eqs.~(\ref{38}) and (\ref{27}) for $\eta=0$ into Eqs.~(\ref{83}) and (\ref{84}), we find the mean occupation numbers for the Maxwell-Boltzmann statistics of particles in the Tsallis unnormalized statistics as
\begin{eqnarray}\label{96}
 \langle n_{\mathbf{p}\sigma} \rangle &=& \sum\limits_{n=0}^{\infty} \frac{\omega^{n}}{n!} \frac{1}{Z^{q}} \frac{1}{\Gamma\left(\frac{q}{q-1}\right)} \int\limits_{0}^{\infty} t^{\frac{1}{q-1}-n} \nonumber \\ &\times&
 e^{-t\left[1-(1-q)\frac{\varepsilon_{\mathbf{p}}-\mu(n+1)}{T}\right]} \nonumber \\ &\times& \left(K_{2}\left(\frac{t(q-1)m}{T} \right)\right)^{n} dt \quad \mathrm{for} \quad q>1
\end{eqnarray}
and
\begin{eqnarray}\label{97}
 \langle n_{\mathbf{p}\sigma} \rangle &=& \sum\limits_{n=0}^{\infty} \frac{(-\omega)^{n}}{n!} \frac{1}{Z^{q}} \Gamma\left(\frac{1}{1-q}\right)  \frac{i}{2\pi} \oint\limits_{C} (-t)^{\frac{1}{q-1}-n} \nonumber \\ &\times& e^{-t\left[1-(1-q)\frac{\varepsilon_{\mathbf{p}}-\mu(n+1)}{T}\right]} \nonumber \\ &\times& \left(K_{2}\left(\frac{t(q-1)m}{T} \right)\right)^{n} dt \quad \mathrm{for} \quad q<1.
\end{eqnarray}
In the ultrarelativistic limit $(m=0)$, Eqs.~(\ref{96}) and (\ref{97}) recover Eq.~(66) of Ref.~\cite{Parvan17}.

Let us find the mean number of particles in the system. Substituting Eqs.~(\ref{38}) and (\ref{44}) into Eqs.~(\ref{89}) and (\ref{90}), we obtain the mean number of particles for the Maxwell-Boltzmann statistics in the Tsallis unnormalized statistics as
\begin{eqnarray}\label{98}
 \langle N \rangle &=& \sum\limits_{n=0}^{\infty} \frac{\omega^{n+1}}{n!} \frac{1}{Z^{q}} \frac{1}{\Gamma\left(\frac{q}{q-1}\right)} \int\limits_{0}^{\infty} t^{\frac{1}{q-1}-n-1} \nonumber \\ &\times&
 e^{-t\left[1+(1-q)\frac{\mu(n+1)}{T}\right]} \nonumber \\ &\times& \left(K_{2}\left(\frac{t(q-1)m}{T} \right)\right)^{n+1} dt \quad \mathrm{for} \quad q>1
\end{eqnarray}
and
\begin{eqnarray}\label{99}
 \langle N \rangle &=&\sum\limits_{n=0}^{\infty} \frac{(-\omega)^{n+1}}{n!} \frac{1}{Z^{q}}  \Gamma\left(\frac{1}{1-q}\right)  \frac{i}{2\pi} \oint\limits_{C} (-t)^{\frac{1}{q-1}-n-1} \nonumber \\ &\times& e^{-t\left[1+(1-q)\frac{\mu(n+1)}{T}\right]} \nonumber \\ &\times& \left(K_{2}\left(\frac{t(q-1)m}{T} \right)\right)^{n+1} dt \quad \mathrm{for} \quad q<1.
\end{eqnarray}
In the ultrarelativistic limit $(m=0)$, Eqs.~(\ref{98}) and (\ref{99}) recover Eq.~(67) of Ref.~\cite{Parvan17}.

Let us find the  mean energy of the system. Substituting Eqs.~(\ref{38}) and (\ref{47}) into Eqs.~(\ref{87}) and (\ref{88}), we obtain the mean energy of the system for the Maxwell-Boltzmann statistics of particles in the Tsallis unnormalized statistics as
\begin{eqnarray}\label{100}
 \langle H \rangle &=& \sum\limits_{n=0}^{\infty} \frac{\omega^{n+1}}{n!} \frac{1}{Z^{q}} \frac{1}{\Gamma\left(\frac{q}{q-1}\right)} \int\limits_{0}^{\infty} t^{\frac{1}{q-1}-n-1} \nonumber \\ &\times&
 e^{-t\left[1+(1-q)\frac{\mu(n+1)}{T}\right]}  \left(K_{2}\left(\frac{t(q-1)m}{T} \right)\right)^{n+1} \nonumber \\
 &\times& \left[m\frac{K_{3}\left(\frac{t(q-1)m}{T} \right)}{K_{2}\left(\frac{t(q-1)m}{T} \right)}- t^{-1}\frac{T}{q-1}  \right] dt \nonumber \\ && \quad \mathrm{for} \quad q>1
\end{eqnarray}
and
\begin{eqnarray}\label{101}
 \langle H \rangle &=& \sum\limits_{n=0}^{\infty} \frac{(-\omega)^{n+1}}{n!} \frac{1}{Z^{q}} \Gamma\left(\frac{1}{1-q}\right)  \frac{i}{2\pi} \oint\limits_{C} (-t)^{\frac{1}{q-1}-n-1} \nonumber \\ &\times& e^{-t\left[1+(1-q)\frac{\mu(n+1)}{T}\right]} \left(K_{2}\left(\frac{t(q-1)m}{T} \right)\right)^{n+1} \nonumber \\
 &\times& \left[m\frac{K_{3}\left(\frac{t(q-1)m}{T} \right)}{K_{2}\left(\frac{t(q-1)m}{T} \right)}- t^{-1}\frac{T}{q-1}  \right] dt \nonumber \\ &&  \quad \mathrm{for} \quad q<1.
\end{eqnarray}
In the ultrarelativistic limit $(m=0)$, Eqs.~(\ref{100}) and (\ref{101}) recover Eq.~(68) of Ref.~\cite{Parvan17}.

Substituting Eqs.~(\ref{96}) and (\ref{97}) into Eq.~(\ref{35}), we obtain Eqs.~(\ref{102}) and (\ref{103}) for the transverse momentum distribution of the Maxwell-Boltzmann statistics of particles in the Tsallis unnormalized statistics.

\subsection{Zeroth term approximation}
Substituting $Z=1$ into Eqs.~(\ref{83}) and (\ref{84}) and retaining only the zeroth term ($n=0$), we obtain
\begin{equation}\label{104}
 \langle n_{\mathbf{p}\sigma} \rangle =  \frac{1}{\Gamma\left(\frac{q}{q-1}\right)} \int\limits_{0}^{\infty} t^{\frac{1}{q-1}} e^{-t} \frac{1}{e^{\beta' (\varepsilon_{\mathbf{p}}-\mu)}+\eta}  dt \quad \mathrm{for} \quad q>1
\end{equation}
and
\begin{eqnarray}\label{105}
 \langle n_{\mathbf{p}\sigma} \rangle &=& \Gamma\left(\frac{1}{1-q}\right)  \frac{i}{2\pi} \oint\limits_{C} (-t)^{\frac{1}{q-1}} e^{-t}  \frac{1}{e^{\beta' (\varepsilon_{\mathbf{p}}-\mu)}+\eta} dt  \nonumber \\ && \mathrm{for} \quad q<1.
\end{eqnarray}
Using Eqs.~(\ref{10}), (\ref{11}) and (\ref{104}), (\ref{105}), we obtain the mean occupation numbers in the zeroth term approximation in the Tsallis unnormalized statistics as
\begin{eqnarray}\label{107}
\langle n_{\mathbf{p}\sigma} \rangle &=& \sum\limits_{k=0}^{\infty} (-\eta)^{k} \left[1+(k+1) (q-1) \frac{\varepsilon_{\mathbf{p}}-\mu}{T} \right]^{\frac{q}{1-q}} \nonumber \\ && \mathrm{for} \quad  \eta=-1,0,1.
\end{eqnarray}
In the ultrarelativistic limit $(m=0)$, Eq.~(\ref{107}) with $\eta=0$ recovers Eq.~(74) of Ref.~\cite{Parvan17}. Substituting Eq.~(\ref{107}) into Eq.~(\ref{85}), we obtain the mean energy of the system in the zeroth term approximation in the Tsallis unnormalized statistics as
\begin{eqnarray}\label{110}
\langle H \rangle &=& \sum\limits_{k=0}^{\infty} (-\eta)^{k} \sum\limits_{\mathbf{p},\sigma} \varepsilon_{\mathbf{p}} \left[1+(k+1) (q-1) \frac{\varepsilon_{\mathbf{p}}-\mu}{T} \right]^{\frac{q}{1-q}} \nonumber \\ && \mathrm{for} \quad \eta=-1,0,1.
\end{eqnarray}
Now substituting Eq.~(\ref{107}) into Eq.~(\ref{86}), we obtain the mean number of particles of the system in the zeroth term approximation in the Tsallis unnormalized statistics as
\begin{eqnarray}\label{113}
\langle N \rangle &=& \sum\limits_{k=0}^{\infty} (-\eta)^{k} \sum\limits_{\mathbf{p},\sigma} \left[1+(k+1) (q-1) \frac{\varepsilon_{\mathbf{p}}-\mu}{T} \right]^{\frac{q}{1-q}} \nonumber \\ && \mathrm{for} \quad \eta=-1,0,1.
\end{eqnarray}

Substituting Eq.~(\ref{107}) into Eq.~(\ref{35}), we obtain Eq.~(\ref{116}) for the transverse momentum distribution of hadrons in the zeroth term approximation in the Tsallis unnormalized statistics.


\end{document}